\documentclass[%
 reprint,
superscriptaddress,
 amsmath,amssymb,
 aps,
]{revtex4-2}

\usepackage[T1]{fontenc}
\usepackage{graphicx}
\usepackage[version=3]{mhchem}
\usepackage{dcolumn}
\usepackage{bm}
\usepackage{siunitx}
\usepackage{placeins}
\usepackage{hyperref}

\usepackage{braket}
\usepackage{algorithm}
\usepackage{algpseudocode}
\usepackage{tabularx}
\usepackage[capitalise]{cleveref}

\newcounter{milestonecount}

\newcommand{\IL}{I_\mathrm{L}}

\newcommand{\VL}{V_\mathrm{L}}
\newcommand{\VR}{V_\mathrm{R}}
\newcommand{\VLM}{V_\mathrm{LM}}
\newcommand{\VRM}{V_\mathrm{RM}}
\newcommand{\VABS}{V_\mathrm{ABS}}

\newcommand{\VABSi}[1]{V_\mathrm{ABS,#1}}

\newcommand{\GLL}{G_\mathrm{LL}}
\newcommand{\GRR}{G_\mathrm{RR}}
\newcommand{\GLR}{G_\mathrm{LR}}
\newcommand{\GRL}{G_\mathrm{RL}}

\makeatletter
\def\@affilnum#1{\textsuperscript{\normalfont#1}}
\makeatother

\begin{document}

\title{Cross-Platform Autonomous Control of Minimal Kitaev Chains}

\author{David~van~Driel}\email{D.vanDriel@tudelft.nl}
\affiliation{QuTech and Kavli Institute of NanoScience, Delft University of Technology, 2600 GA Delft, The Netherlands}

\author{Rouven~Koch}
\affiliation{QuTech and Kavli Institute of NanoScience, Delft University of Technology, 2600 GA Delft, The Netherlands}
\affiliation{Department of Applied Physics, Aalto University, 02150 Espoo, Finland}

\author{Vincent~P.~M.~Sietses}
\author{Sebastiaan~L.~D.~ten~Haaf}
\author{Chun-Xiao~Liu}
\affiliation{QuTech and Kavli Institute of NanoScience, Delft University of Technology, 2600 GA Delft, The Netherlands}
\author{Francesco~Zatelli}
\author{Bart~Roovers}
\author{Alberto~Bordin}
\author{Nick~van~Loo}
\author{Guanzhong~Wang}
\author{Jan~Cornelis~Wolff}
\author{Grzegorz~P.~Mazur}
\author{Tom~Dvir}
\author{Ivan~Kulesh}
\author{Qingzhen~Wang}
\author{A.~Mert~Bozkurt}
\affiliation{QuTech and Kavli Institute of NanoScience, Delft University of Technology, 2600 GA Delft, The Netherlands}

\author{Sasa~Gazibegovic}
\author{Ghada~Badawy}
\author{Erik~P.~A.~M.~Bakkers}
\affiliation{Department of Applied Physics, Eindhoven University of Technology, 5600 MB Eindhoven, The Netherlands}

\author{Michael~Wimmer}
\author{Srijit~Goswami}
\affiliation{QuTech and Kavli Institute of NanoScience, Delft University of Technology, 2600 GA Delft, The Netherlands}

\author{Jose~L.~Lado}
\affiliation{Department of Applied Physics, Aalto University, 02150 Espoo, Finland}

\author{Leo~P.~Kouwenhoven}
\affiliation{QuTech and Kavli Institute of NanoScience, Delft University of Technology, 2600 GA Delft, The Netherlands}

\author{Eliska Greplova}\email{e.greplova@tudelft.nl}
\affiliation{QuTech and Kavli Institute of NanoScience, Delft University of Technology, 2600 GA Delft, The Netherlands}

\date{\today}

\begin{abstract}

Contemporary quantum devices are reaching new limits in size and complexity, allowing for the experimental exploration of emergent quantum modes. However, this increased complexity introduces significant challenges in device tuning and control. Here, we demonstrate autonomous tuning of emergent Poor Man's Majorana zero modes in a minimal realization of a Kitaev chain. 
We achieve this task using cross-platform transfer learning. First, we train a tuning model on a theory model. Next, we retrain it using a Kitaev chain realization in a two-dimensional electron gas. Finally, we apply this model to tune a Kitaev chain realized in quantum dots coupled through a semiconductor-superconductor section in a one-dimensional nanowire.
Utilizing a convolutional neural network, we predict the tunneling and Cooper pair splitting rates from differential conductance measurements, employing these predictions to adjust the electrochemical potential to a Poor Man's Majorana sweet spot. The algorithm successfully converges to an immediate vicinity of a sweet spot (within $\pm \SI{1.5}{\milli\volt}$ in 67.6\% of attempts and within $\pm \SI{4.5}{\milli\volt}$ in 80.9\% of cases), typically finding a sweet spot in 45 minutes or less. This advancement is a stepping stone towards autonomous tuning of emergent modes in interacting systems, and towards foundational tuning machine learning models that can be deployed across a range of experimental platforms.

\end{abstract}

\maketitle

\section*{Introduction}

Mesoscopic quantum devices are promising for both quantum computing \cite{philips2022universal, kim2023evidence, mi2021information} and quantum simulation \cite{van2021quantum, rosenberg2023dynamics, shtanko2023uncovering, jouanny2024band,splitthoff2024gate} due to the flexibility and scalability in the design and fabrication. However, a challenge in terms of control and tuning emerges as these devices reach larger sizes. In other words, the scalability of solid state quantum devices comes with the price of complex quantum control requirements. There are two main lines of inquiry towards solving these challenges either based on (i) the development of effective models that can be embedded into the experimental control protocols \cite{klimov2024optimizing, bengtsson2024model}, or (ii) data-driven machine learning (ML) approaches that are trained to extract key features and make predictions directly on the measured data \cite{dawid2022modern, durrer2020automated, schuff2024fully, zwolak2020autotuning, lienhard2022deep, barrett2023learning, kalantre2019machine, lennon2019efficiently, nguyen2021deep, darulova2020autonomous, czischek2021miniaturizing, Severin2024,koch2023adversarial}. The latter technique has shown success in cases where precise theoretical modelling is challenging or when the algorithmic application of effective model is prohibitive due to its complexity. In instances, where there is a sufficient amount of training data available, ML models show impressive generalization with respect to noise and, once trained, rapid evaluation speed that allows for embedding into the existing experimental workflows \cite{zwolak2023colloquium}.

In particular, ML techniques have been widely deployed for tasks pertaining to the operation of semiconductor quantum dots \cite{zwolak2023colloquium}. In this context ML methods have been applied with promising results that simplify device operation and reach comparable accuracy to experienced human operators \cite{moon2020machine}. The key challenges in the field can be grouped into two broader categories \cite{zwolak2024data}: (i) building foundational models that generalize well to a large set of platforms and devices, and, (ii) demonstrating the utility of machine learning approaches beyond smaller-scale devices with relatively simple charge stability diagrams and a comparably low number of parameters.

In the present work, we design, benchmark and experimentally demonstrate a ML-based method that addresses both these challenges simultaneously: we devise an autonomous tune-up protocol that achieves the emergence of Poor Man's Majorana modes (PMMs), and we deploy the protocol without using data from the experimental platform we wish to tune. Specifically, we concentrate on the experimental realization of PMMs that are predicted to manifest as emergent quasiparticles in a one-dimensional Kitaev chain~\cite{kitaev2001unpaired, dvir2023realization}.
Kitaev chains have been modeled in various forms, including both spinless and spinful variants~\cite{Kitaev2001, Sau2012, Beenakker2013, Pan2023, Liu2024, Tsintzis2022, Luethi2024, Tsintzis2024, Samuelson2024, Souto2023}.
By tuning up complex emergent modes, we take a step beyond what has been established in autonomous state preparation previously. Moreover, we achieve this tuning in a cross-platform transfer learning setting. Specifically, we consider three objects during testing and training of our ML algorithm: a theoretical model of the Kitaev chain, a Kitaev chain experimental realization in chain of quantum dots coupled to superconductor in a one-dimensional nanowire (Device A), a Kitaev chain experimental realization based on two-dimensional electron gas (Device B). We train the tuning algorithm on theoretical model, re-train on a small number of data acquired from Device B and then apply and evaluate its performance on Device A.

Our method relies on conductance measurements of charge stability diagrams and works on any platform that supports them. The PMMs in a Kitaev chain originate from two distinct physical processes. A convolutional neural network (CNN) identifies the dominant process. At the voltage configuration where these two processes are equal, the PMMs emerge and we refer to this point as the sweet spot.
The output of this initial classification is then used as an input for a gradient descent algorithm that minimizes the distance to the sweet spot by adjusting the device's gate voltages. We find that our algorithm is able to successfully converge to the vicinity of the sweet spot in the voltage space where the PMMs are predicted to emerge. Specifically, 67.6\% of tuning runs converge within $\pm \SI{1.5}{\milli\volt}$, and 80.9\% converge within $\pm \SI{4.5}{\milli\volt}$ of a known sweet spot.

This work is organized as follows: First we explain the details and characterization of the two-dot Kitaev chain we aim to tune. Then we describe all the stages of the tuning algorithm: the theory model, the validation and re-training of the model on Device B, and finally its application to Device A. Then we summarize the ML algorithm performance, and assess and summarize our findings in the discussion and conclusion sections. We refer readers to a complementary submission by Benestad et al. that simulates a tuning algorithm based on additional sensor quantum dots coupled to a minimal Kitaev chain~\cite{benestad2024machine}.

\section*{Device A characterization}

\begin{figure*}[ht!]
    \centering
    \includegraphics[width=\linewidth]{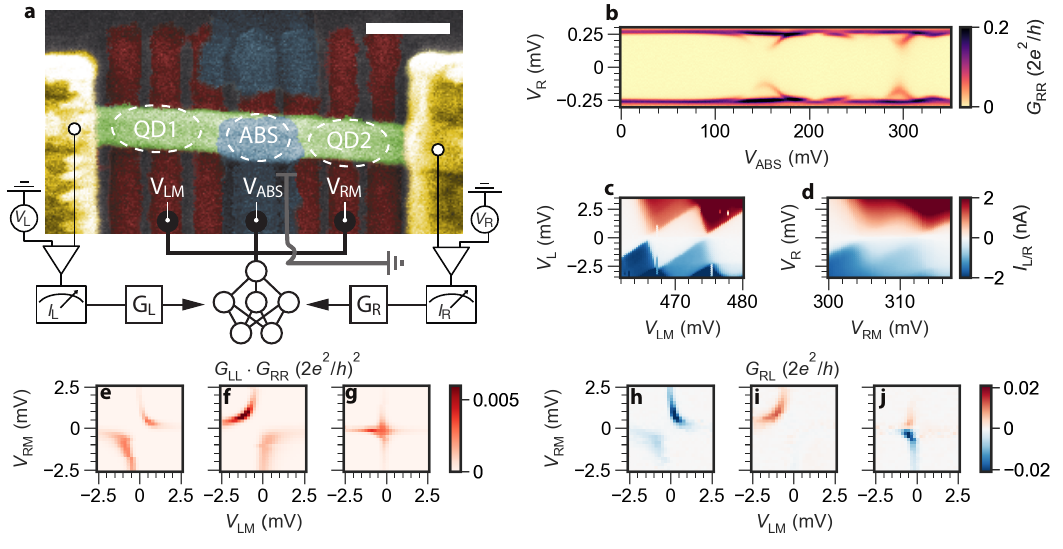}
    \caption{\textbf{Device A set-up and characterization.} 
    \textbf{a.} False color SEM micrograph of our device, and the circuit. An InSb nanowire (green) is contacted by a grounded Al shell (blue) and two normal leads (yellow). The nanowire is placed on bottom gates (red). The relevant gate voltages are indicated by text. Both normal leads can be voltage biased independently with respect to the grounded Al.
    \textbf{b.} Local differential conductance measured from the right lead, $\GRR$, for varying hybrid plunger gate voltage, $\VABS$, and $\VR$ at $B = \SI{0}{\tesla}$. 
    \textbf{c.} The current measured at the left lead, $I_\mathrm{L}$, for varying plunger gate voltage of the left quantum dot (QD), $\VLM$, and bias voltage applied to the left lead, $\VL$. 
    \textbf{d.} The current measured at the right lead, $I_\mathrm{R}$, for varying plunger gate voltage of the right QD, $\VRM$, and bias voltage applied to the right lead, $\VR$.
    \textbf{e.-g.} The product of the left and right differential conductance, $\GLL \cdot \GRR$, for varying $\VLM$ and $\VRM$ at $\VL=\VR=\SI{0}{\volt}$. The measurements were performed at $\VABS$ values of \SI{85}{\milli\volt}, \SI{104}{\milli\volt} and \SI{115}{\milli\volt} respectively.
    \textbf{h.-j.} Non-local conductance, $\GRL = dI_\mathrm{R}/dV_\mathrm{L}$ for varying $\VLM$ and $\VRM$. The data was measured simultaneously with panels e-g.
    }\label{fig: 1}
\end{figure*}

In this section, we describe Device A, on which we demonstrate the tuning algorithm. The device is a semiconductor nanowire designed to be a minimal Kitaev chain, consisting of two quantum dots (QDs) coupled by a semiconductor-superconductor segment, as in the first experimental Kitaev chain realization~\cite{dvir2023realization}. Minimal Kitaev chains are predicted to host non-Abelian statistics, enabling quantum information operations such as fusion, braiding, and qubit formation~\cite{liu2023fusion, boross_braiding-based_2023, tsintzis2024majorana}.
We show a fake-color SEM micrograph of device A in \cref{fig: 1}a. The nanowire (green) is deposited on bottom finger gates (red), and contacted by normal Cr/Au (yellow) contacts at its ends. 
The middle section of the nanowire, the hybrid, is contacted by a grounded Al shell (blue). Further details pertaining to the device fabrication can be found in Refs.~\cite{heedt_shadow-wall_2021, van_driel_spin-filtered_2022}.

The left and right normal leads can be biased with respect to the grounded Al with voltages $\VL$ and $\VR$ respectively. The current running through the left lead ($I_\mathrm{L}$) and through the right lead ($I_\mathrm{R}$) can be measured independently. Unless explicitly indicated, we do not apply an external magnetic field.
Using standard lock-in detection, we obtain the local and non-local differential conductances~\cite{menard2020conductance}:
\begin{equation*}
	\begin{aligned}
		&\GLL = dI_\mathrm{L}/dV_\mathrm{L}, &\GRR = dI_\mathrm{R}/dV_\mathrm{R}\\
            &\GRL = dI_\mathrm{R}/dV_\mathrm{L}, &\GLR = dI_\mathrm{L}/dV_\mathrm{R}\\
        \end{aligned}
\end{equation*}
Measurements of the local conductances, $\GLL$ and $\GRR$, for varying $\VLM$ and $\VRM$ are used as an input for a convolutional neural network (CNN) that predicts the rates of elastic co-tunneling (ECT) and crossed Andreev reflection (CAR) between the two QDs (see \cref{app:architecture} for details on the CNN architecture). The PMMs appears when the ECT and CAR rates are equal \cite{sau2012realizing, liu2022tunable, dvir2023realization}. The prediction from the CNN is then used as input for a gradient descent algorithm that adjusts $\VABS$ until the rates are equal, marking a sweet spot. While we could have equally used $\GRL$ and $\GLR$ as input for the CNN, measurements of non-local conductance require a multi-terminal set-up that does not scale well for longer Kitaev chains. 

First, we characterize the spectrum of the hybrid segment by accumulating electrons in the InSb nanowire, and defining tunnel barriers using the two gates directly next to the hybrid segment. 
In \cref{fig: 1}b, we show $\GRR$ for varying $\VR$ and hybrid plunger gate voltage, $\VABS$. A hard superconducting gap is seen over the entire gate voltage range. Multiple discrete states are observed in this range, whose gate dependence is typical for Andreev bound states (ABSs).
Then, we change the tunnel gate voltages to define quantum dots (QDs) to the left and right of the hybrid. 
We control their electrochemical potentials by varying the plunger gate voltages $\VLM$ and $\VRM$. 
We characterize the left QD by varying $\VLM$ and $\VL$ while measuring $\IL$  as shown in \cref{fig: 1}c. 
For small bias voltage the current is fully suppressed due to the Coulomb blockade, and as the bias increases we see current appearing in the form of Coulomb diamonds. Additionally, we notice that a gap has opened around the Fermi level due to the superconducting Al. 
We also perform an analogous measurement of the right QD as shown in \cref{fig: 1}d. 
Here, the Coulomb diamonds are more difficult to resolve than in \cref{fig: 1}c, which we attribute to a stronger tunnel-coupling to the superconductor \cite{jellinggaard2016tuning}.

After defining the QDs, we set $\VABS=\SI{85}{\milli\volt}$ and measure charge stability diagrams (CSDs). 
\cref{fig: 1}e, shows the correlated local conductance, $\GLL \GRR$, for varying $\VLM$ and $\VRM$. This correlated conductance is only finite when transport processes involve both QDs. Therefore, it filters out features that are local to only one QD, such as Andreev reflection. We see an avoided crossing that indicates hybridization between the left and right QD levels. \cref{fig: 1}h shows the non-local conductance, $\GRL$, that was measured simultaneously. 
The negative sign of $\GRL$ indicates that the ECT rate is greater than the CAR rate, which is confirmed by the anti-diagonal avoided crossing \cite{dvir2023realization}.
\cref{fig: 1}f, i show $\GLL \GRR$ and $\GRL$ respectively, for varying $\VLM$ and $\VRM$ at $\VABS=\SI{104}{\milli\volt}$. 
Here, the avoided crossing is now diagonal and the sign of $\GRL$ is positive, indicating that CAR dominates over ECT. 
\cref{fig: 1}g, j show $\GLL \GRR$ and $\GRL$ at $\VABS=\SI{115}{\milli\volt}$. 
There is no avoided crossing in the CSD anymore, and $\GRL$ has both negative and positive values, which is characteristic of the sweet spot where the ECT and CAR rates are equal. 

\section*{Tuning Algorithm}
\begin{figure}[ht!]
    \centering
    \includegraphics[width=\linewidth]{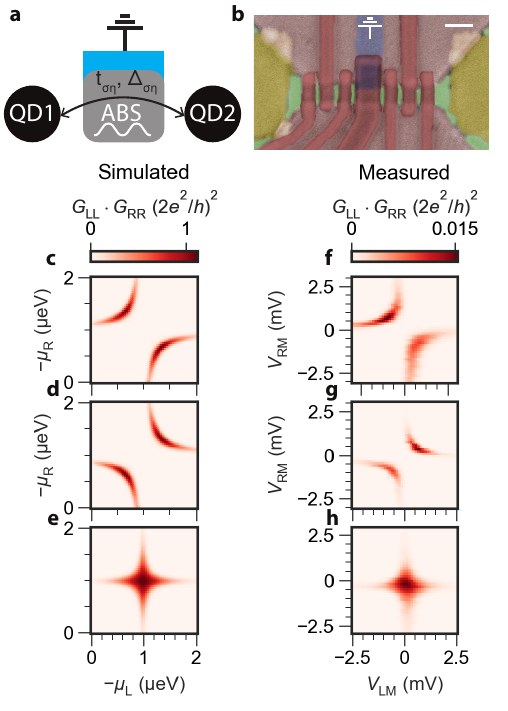}
    \caption{\textbf{Device B charge stability diagrams and simulations.} 
    \textbf{a.} Schematic overview of the model system. Two quantum dots (QDs) are coupled by an Andreev bound state (ABS) in a semiconductor (gray) coupled to a grounded superconductor (blue).
    \textbf{b.} False color SEM image of device B.  A 1D channel is defined in the InAsSb (green) using a top and bottom depletion gate (light red). The InAsSb is contacted in the middle by a grounded Al strip (blue). QDs are defined the left and right of this hybrid segment using finger gates (darker red). The InAsSb channel is contacted from the sides by two Ti/Au leads (yellow). Scale bar is \SI{200}{\nano\meter}.
    \textbf{c-e.} Simulated conductance, $\GLL \cdot \GRR$ at zero energy, for varying chemical potential of the left and right quantum dots, $\mu_\mathrm{L}$ and $\mu_\mathrm{R}$. These were generated for $(t-\Delta)/(t+\Delta) = \{-1/3,\ 1/3, \ 0 \}$ respectively. 
    \textbf{f-h.} Measured $\GLL \cdot \GRR$, for varying $\VLM$ and $\VRM$ at $\VL=\VR=\SI{0}{\volt}$ for device B. The CNN predicts $(t-\Delta)/(t+\Delta) = \{-0.31, \ 0.31,\ 0.01 \}$ for these respectively.
    }\label{fig: 2}
\end{figure}
\subsection*{Theory Model}
To train a CNN to predict the CAR and ECT rates from CSDs, we construct an effective model of the system to simulate transport data. We model the device in \cref{fig: 1}a as a 3-site system, where the semiconductor-superconductor hybrid segment is treated as an individual site, as shown in \cref{fig: 2}a. The couplings between the two QDs are mediated by ABSs in the hybrid segment.
A more detailed description of the theory model and Hamiltonian can be found in \cref{app:training}. We can write the effective interaction Hamiltonian for the QDs in the occupation basis as:
\begin{equation}
    H_{\mathrm{T}} = \sum_{\sigma,\eta=\uparrow,\downarrow} \left( t_{\sigma\eta} c_{DL\sigma}^{\dagger} c_{DR\eta} + \Delta_{\sigma\eta} c_{DL\sigma}^{\dagger} c_{D R \eta}^{\dagger} \right) + h.c. , \label{eq:Heff}
\end{equation} 
The operators $c_{DL/DR}^\dagger$ ($c_{DL/DR}$) create (annihilate) an electron is each of the QDs. $t_{\sigma\eta}$ and $\Delta_{\sigma\eta}$ are the ECT and CAR couplings between the two QDs with corresponding spins in the left and right QD, $\sigma$ and $\eta$, respectively. Coupling to the ABS results in an induced gap on each QD, leading us to model them as Yu-Shiba-Rusinov (YSR) states~\cite{YULUH1965,Shiba1968,rusinov1969theory}. We add the QDs' electrochemical potential and charging energy to Eq.~\eqref{eq:Heff} to obtain the full Hamiltonian (see \cref{eq:fullH} for details). We then perform a Bogoliubov transformation to account for the particle-hole symmetry on the QDs, and write an effective Hamiltonian for two coupled YSR states in the limit of infinite Zeeman splitting as:
\begin{equation}
\begin{split}
    H_{\mathrm{eff}}& =  \delta \varepsilon_L f_L^\dagger f_L + \delta \varepsilon_R f_R^\dagger f_R \\
    &+ t \left(f_L^\dagger f_R + f_R^\dagger f_L\right) + \Delta \left(f_L^\dagger f_R^\dagger + f_L f_R\right).
\end{split} \label{eq: effectiveH}
\end{equation}
The operators $f_{L/R}^\dagger$ ($f_{L/R}$) create (annihilate) a quasiparticle in each QD, that is a superposition of electron and hole components.
Here, $\delta \varepsilon_{L/R}$ are the energies of the YSR states and $t$ and $\Delta$ are the generalized effective couplings for the odd-, and even-parity states respectively. While $t$ and $\Delta$ correspond to ECT and CAR for normal QDs, this is no longer the case for YSR states. Due to local Andreev reflection, ECT can now also couple the global even-parity states, and CAR the odd states (see \cref{app:training} for details). We note that \cref{eq: effectiveH} is a spinless Hamiltonian when expressed in terms of the YSR states for which localized Majorana zero modes (MZMs) will emerge at $t=\Delta$. On the other hand, at zero field, the odd states are spin-degenerate and Eq.~\eqref{eq: effectiveH} is no longer valid. However, the system has a sweet spot at $t=\tilde{\Delta} (B=0) = \sqrt{2}\Delta$. Here, the presence of high charging energy in the quantum dots leads to the emergence of Majorana Kramers pairs that are protected against local perturbations but have an additional degeneracy (see Ref.~\cite{bozkurt2024} for details). The CSD in the sweet spot is identical for zero and infinite Zeeman energy. As a result, we can train the CNN on the spinless Hamiltonian in Eq.~\eqref{eq: effectiveH}, and recognize sweet spots in both the absence and presence of a magnetic field. We shall refer to the renormalized and field-dependent parameter $\tilde{\Delta}(B)$ as $\Delta$, to ensure that $t=\Delta$ corresponds to a sweet spot for all magnetic field values.
We have further validated our approach using a spinful Hamiltonian model (Appendix~\ref{app:spinful hamiltonian}), confirming that the main conclusions remain unchanged.

\cref{eq: effectiveH} allows us to compute CSDs using the Mahaux-Weidenmuller formula and the scattering matrix formalism~\cite{Christiansen2009} (see. App.~\ref{app:training} for more details). 
We generate CSDs for varying values of $t$ and $\Delta$, and show three examples in \cref{fig: 2}c-e. We label each generated CSD with $(t-\Delta)/(t+\Delta)$ and train the CNN to predict these labels on 6000 examples (see \cref{app:training} for details on the training set). We use this ratio for labeling, as it is a dimensionless quantity and scales with the interaction strength $t+\Delta$ of the QDs. In principle, we could have used $\sqrt{8 \lvert t^2-\Delta^2 \rvert }$, which is directly proportional to the distance between the hyperbolas of the avoided crossing in the CSDs~\cite{ten2024two}. However, there is a scaling factor that depends on the measurement resolution, the lever arm of the gates and the range of $\VLM$ and $\VRM$. To eliminate the necessity to manually scale an absolute quantity, we choose the dimensionless ratio, $(t-\Delta)/(t+\Delta)$.

\subsection*{Model Validation and Re-training on Device B}
To develop a cross-platform tuning algorithm, we require the CNN predictions to generalize across different device types. 
Before applying the CNN to device A, we first perform measurements on device B, which was fabricated on an InAsSb-Al two-dimensional electron gas (2DEG), shown in \cref{fig: 2}b (see also Ref.~\cite{ten2024two}). Here, a quasi 1-D channel is defined by two large depletion gates. Gate-defined QDs are created on the left and right of a region proximitized by a thin Al strip (blue), using fabrication methods detailed in \cite{Moehle2022}. The same measurement set-up is used as for device A.
In \cref{fig: 2}f-h, we show three example CSDs measured at different $\VABS$ values using device B. These CSDs are used as inputs for our trained CNN, yielding predictions for $(t-\Delta)/(t+\Delta)$ of ${-0.31, \ 0.31,\ 0.01 }$, respectively. The experimental CSDs closely resemble the generated ones in \cref{fig: 2}c-e, which we manually matched with the experimental ones by setting: $(t-\Delta)/(t+\Delta) = \{-1/3,\ 1/3, \ 0 \}$ respectively. 

As $t$ and $\Delta$ represent the ECT and CAR rates, validating the CNN predictions requires determining these rates quantitatively. We extract them by performing bias spectroscopy at the $\VLM, \VRM$ values corresponding to the center of the (avoided) crossing of the CSDs. Typically, each spectrum consists of four conductance peaks~\cite{zatelli2024robust}:
two inner peaks separated by $\lvert t-\Delta \rvert$ and two outer peaks separated by $t+\Delta$ (see \cref{app:label} for details on the labeling procedure). The ratio of the peak distances can then be used to calculate $(t-\Delta)/(t+\Delta)$, which is combined with the corresponding CSD to constitute one labeled dataset.

\cref{fig: 3}a compares the $(t-\Delta)/(t+\Delta)$ values obtained experimentally with those predicted by the CNN for a range of $\VABS$ values. The CNN predictions (red markers) generally match the values extracted from bias spectroscopy (black markers), with small deviations around $(t-\Delta)/(t+\Delta)=0$. We note that the labeled data implies that $(t-\Delta)/(t+\Delta)=0$ for a range of $\VABS$ values. However, only the CSD shown in \cref{fig: 3}d indicates a clear sweet spot. We attribute this apparent plateau of $(t-\Delta)/(t+\Delta)=0$ to the influence of multiple, overlapping ABSs in the hybrid section of device B (see extended data Fig.~1e of \cite{tenHaaf2024}). In addition, our labeling procedure is limited by tunnel broadening and temperature, see \cref{app:label}.
We stress that the CNN has not been trained on any experimental data for these predictions. To improve the accuracy of the CNN, we retrain it on a part of the experimental data that we manually labeled. For retraining, we use 51 labeled CSDs that were randomly sampled from \cref{fig: 3}a, which is significantly fewer than the 6000 theory datasets initially used to train the CNN (see~\cref{app:training} for details). As labeling experimental data is intensive and time-consuming, it is crucial to minimize the number of labelled CSDs required to retrain the CNN. 





The orange markers in \cref{fig: 3}b show the predictions of the retrained CNN applied to a test set of 24 previously unseen experimental CSDs. 
We observe that the predicted values of $(t-\Delta)/(t+\Delta)$ are closer to the values extracted from spectroscopy compared to the predictions of the CNN that was trained on the simulated data only.
In \cref{fig: 3}c-e we show the CSDs measured for device B for $\VABS$ values indicated by the colored marks in panel b. We see that the avoided crossings in panels c. and e. have reversed direction, which agrees with the predicted sign change of $(t-\Delta)/(t+\Delta)$ shown in \cref{fig: 3}b. The Coulomb resonances form a cross in \cref{fig: 3}d, which is consistent with the predicted $(t-\Delta)/(t+\Delta)\approx0$ and indicates a sweet spot. 

\subsection*{Transfer Learning: Application of the algorithm on Device A}

\begin{figure}[ht!]
    \centering
    \includegraphics[width=\linewidth]{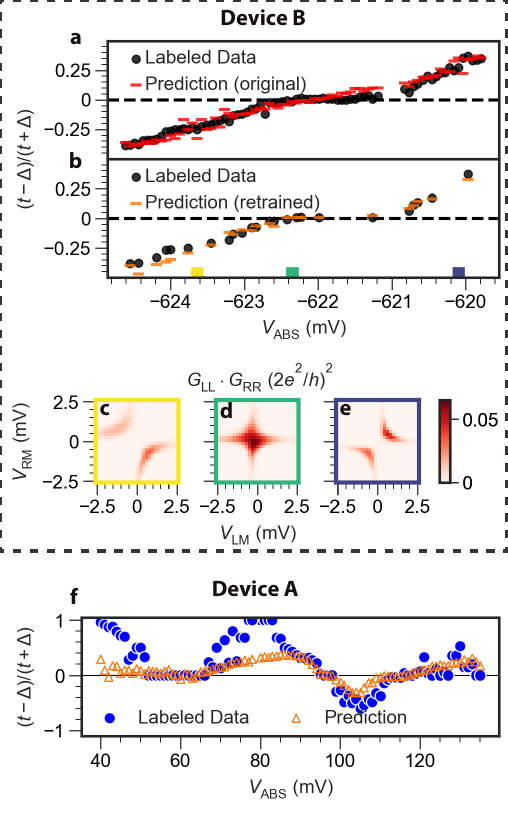}
    \caption{\textbf{Convolutional neural network (CNN) retraining and predictions.} 
    \textbf{a.} The weighted ratio of elastic co-tunneling ($t$) and crossed-Andreev reflection ($\Delta$) rates, $(t-\Delta)/(t+\Delta)$, for varying $\VABS$ for device B. The black markers indicate experimentally-labeled data and the red markers are the CNN predictions.
    \textbf{b.} $(t-\Delta)/(t+\Delta)$ of a subset of test data of panel a after retraining the CNN on experimental data. The black markers indicate experimentally labeled data and the orange markers are the predictions of the retrained CNN.
    \textbf{c.-e.} The product of the left and right differential conductance, $\GLL \cdot \GRR$, for varying $\VLM$ and $\VRM$ at $\VL=\VR=\SI{0}{\volt}$ for device B. The measurements were performed at the $\VABS$ values indicated by colored marks in panel b.
    \textbf{f.} $(t-\Delta)/(t+\Delta)$ for varying $\VABS$ for device A. The blue markers indicate experimentally labeled data, and the orange markers are the retrained CNN predictions. 
    }\label{fig: 3}
\end{figure}

Before discussing the CNN's performance on Device A, it is crucial to note that we only retrained it on experimental data from Device B. The ability of machine learning algorithms to generalize is vital, making them highly effective for analyzing noisy experimental datasets. 
In \cref{fig: 3}a, we assessed the performance of a CNN that was trained on a theoretical model only, which may not fully capture the nuances of experimental data. We now consider a different type of generalization, namely the machine learning model's ability to adapt from one device to another. Eliminating the need for device-specific retraining offers a significant advantage in time.


We now apply the CNN that was trained on theory simulations and re-trained on all 75 experimental datasets from device B (2DEG) to device A (nanowire) that is shown in \cref{fig: 1}a. The blue markers in \cref{fig: 3}f show the $(t-\Delta)/(t+\Delta)$ values that we have extracted experimentally from device A using the same method as for device B. 
The orange markers show the corresponding predictions of the CNN. To compensate for differences in the conductance of the CSD measurements,
we regularize the CNN predictions (see \cref{app:regularization} for details). We see excellent agreement with the experimental data for small $(t-\Delta)/(t+\Delta)$. 
Most importantly, there is good agreement for the values of $\VABS$ at which the $(t-\Delta)/(t+\Delta)$ ratio changes sign, indicating a sweet spot. 
Around $\VABS \approx \SI{50}{\milli\volt}$, the CNN predicts values close to 0, which is not in agreement with the labeled data. 
We attribute this to the left QD interacting strongly with the ABS, which exceeds the simplified model used for generating training data (see \cref{app:restraints} for details). 

For $\VABS$ values close to $\SI{40}{\milli\volt}$ and $\SI{80}{\milli\volt}$, the labeled data indicates $(t-\Delta)/(t+\Delta)\approx1$, while the CNN predictions are considerably lower. Here, $\Delta \ll t$, which makes it challenging to resolve the pairs of conductance peaks at $(t-\Delta)$ and $(t+\Delta)$ in bias spectroscopy, leading to a ratio of $1$ (see \cref{app:label} for details). As the CNN predictions do not depend on peak spacing or measurement resolution, it predicts lower values of $(t-\Delta)/(t+\Delta)$ compared to the labelled data.
The CNN predictions also mildly differ from the labeled data around $\VABS \approx \SI{130}{\milli\volt}$ as well. We attribute this to a low interaction strength, $(t+\Delta)$, which causes the ratio $(t-\Delta)/(t+\Delta)$ to become large. 
As the CNN was trained and retrained predominantly on strongly interacting QDs, it generalizes less well to the device with low $(t+\Delta)$. We conclude that the CNN can correctly identify $t \approx \Delta$ when the QD does not strongly hybridize with the ABS, and when the QD-QD interaction strength is at least larger than the measurement resolution.




\subsection*{Gradient Descent Voltage Optimization}
\begin{figure}[ht!]
    \centering
    \includegraphics[width=\linewidth]{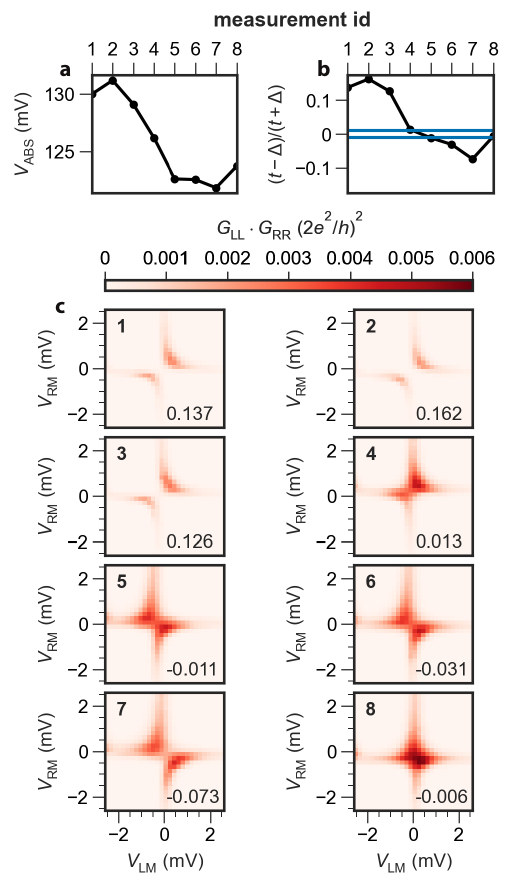}\caption{\textbf{Example of an automated tuning run of device A.} 
    \textbf{a.} The hybrid plunger gate voltage, $\VABS$, for each measurement.
    \textbf{b.} Prediction of the weighted ratio of elastic co-tunneling and crossed-Andreev reflection, $(t-\Delta)/(t+\Delta)$, for each measurement. The horizontal, blue lines indicate the tolerance set before the run.
    \textbf{c.} $\GLL \cdot \GRR$, for varying $\VLM$ and $\VRM$ for subsequent measurements. The measurement ids are indicated by the bold text in the top left corner. The predicted value of $(t-\Delta)/(t+\Delta)$ is indicated in the bottom right corner.
    }\label{fig: 4}
\end{figure}
Each CNN prediction of $(t-\Delta)/(t+\Delta)$ is incorporated into a gradient descent algorithm that sets a new value of the voltage $\VABS$. The algorithm minimizes the cost function $f(\VABS) = \lvert(t-\Delta)/(t+\Delta)\rvert$ until it reaches a value below a set tolerance. In each step, $n$, the algorithm computes the following quantities~\cite{bottou2012stochastic}:
\begin{equation}
    \begin{aligned}
        \braket{g}_n &= \frac{1}{N}\sum_{i = n-N}^n \frac{f(\VABS)_i - f(\VABS)_{i-1}}{\VABS^i - \VABS^{i-1}}, \\
        v^\mathrm{n+1} &= k v^\mathrm{n} + (1-k) \braket{g}_n, \\
        \VABS^{n+1}  &= \VABS^{n}  - \eta v^\mathrm{n+1}.
    \end{aligned} \label{eq:GD}
\end{equation}
Here, $\braket{g}_n$ is the gradient of the objective function which is computed after the $n$th measurement, averaged over the past $N$ measurements. 
The velocity $v^\mathrm{n+1}$ is a mixture of the previous velocity $v^\mathrm{n}$, and gradient $\braket{g}_n$, weighted by the momentum, $k$. 
The change in $\VABS$ is then determined by $v^\mathrm{n+1}$, scaled by the learning rate $\eta$. 
These steps are repeated until $f(\VABS) <\tau $, where $\tau$ is the tolerance set for each run. 
We note that in the first step of the algorithm, $\VABS$ is changed by a pseudo-random number drawn from a Gaussian distribution with a standard deviation of $1$, and an amplitude of $\SI{1.2}{\milli\volt}$. This first step is on the order of typical $\VABS$ changes made by the algorithm.
The minimal change of $\VABS$ is set by the $\SI{60}{\micro\volt}$ resolution of our digital-analog converters. 
We limit the maximal change to $\SI{5}{\milli\volt}$, to prevent large shifts of the QD levels due to cross-capacitance. See \cref{app:gradient} for details on the algorithm, the choices of $\tau$ and $\eta$, and the algorithm pseudocode. 

We now apply the automated tuning algorithm to device A.
\cref{fig: 4}a shows the gate voltages, $\VABS$, set by the algorithm for a run that converged below tolerance in 8 steps. 
\cref{fig: 4}b shows the values of $(t-\Delta)/(t+\Delta)$ predicted by the CNN. Here, the blue horizontal lines indicate the ideal tolerance, $\tau = 0.01$, set before the run. 
We show the CSDs for each step in \cref{fig: 4}c, and see that the direction of the avoided crossing reverses between steps 4 and 5. 
This indicates that $(t-\Delta)/(t+\Delta)$ changes sign, which is confirmed by the CNN predictions in panel b. 
Due to the momentum term $k$ in the gradient descent algorithm (\cref{eq:GD}), the algorithm first proceeds to lower $\VABS$ values before the velocity term changes sign. 
The CNN converges below the tolerance at measurement 8, which we can confirm as a sweet spot by identifying a cross in panel \cref{fig: 4}j. 
In this run, the automated tuning was able to correctly identify the sweet spot, as well as vary $\VABS$ to find it. We show an example of a run that did not converge successfully in \cref{app:unconverged example}.

\section*{Tuning Algorithm Performance}
\subsection*{Zero Field Algorithm Performance}

\begin{figure*}[ht!]
    \centering
    \includegraphics[width=\linewidth]{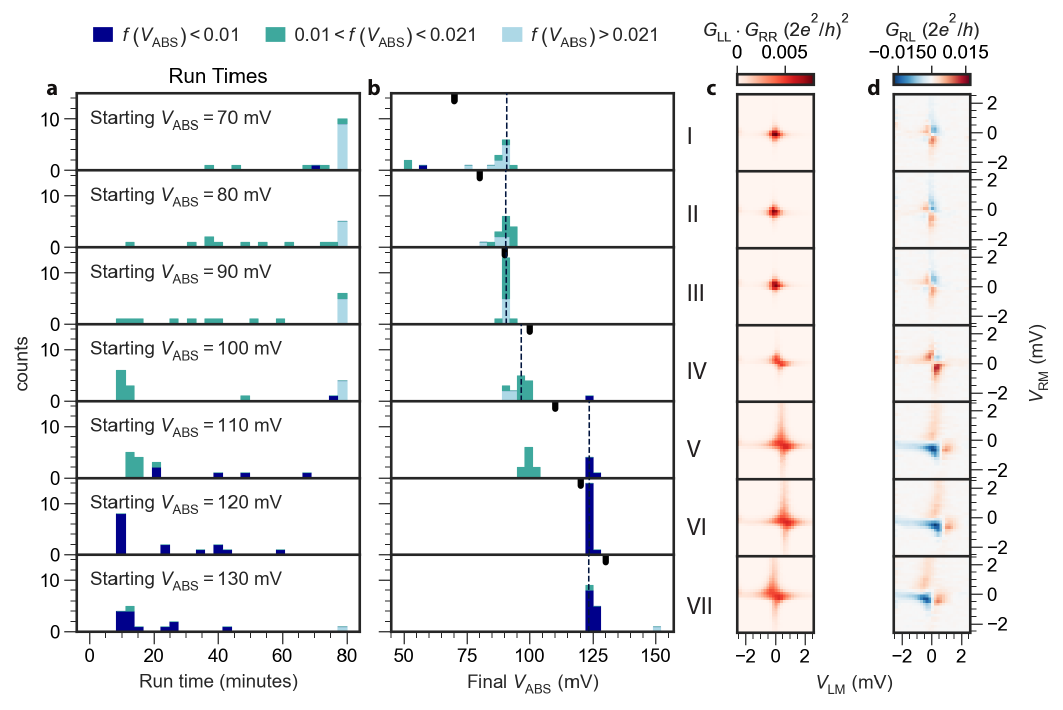}
    \caption{\textbf{Run times and $\VABS$ values at which the algorithm converges for zero applied magnetic field.} We set a learning rate $\eta = \num{2e-4}$, a momentum $k=0.5$ and a tolerance $\tau = \num{1e-2}$.
    \textbf{a.} Histograms of the run time of the gradient descent algorithm, for different starting $\VABS$ values. The dark blue bars indicate which runs converged with an objective function $f(\VABS) = \lvert(t-\Delta)/(t+\Delta)\rvert<0.01$. The sea green bars indicate convergence with $0.01<f(\VABS)<0.021$. Light blue bars indicate runs that did not converge with $f(\VABS)<0.021$ within 25 iterations.
    \textbf{b.} Histograms of the $\VABS$ values for which the algorithm found the lowest $f(\VABS)$, for different starting $\VABS$ values. The black marks at the top of the figures indicate the starting $\VABS$ value. The dashed, vertical line indicates the $\VABS$ value closest to the greatest histogram peak with the lowest $f(\VABS)$.
    \textbf{c.} $\GLL \cdot \GRR$, for varying $\VLM$ and $\VRM$, measured at the $\VABS$ value indicated by the dashed, vertical line in panel b.
    \textbf{d.} $\GRL = dI_\mathrm{R}/dV_\mathrm{L}$ for varying $\VLM$ and $\VRM$, for different starting $\VABS$ values. The data was measured simultaneously with $\GLL \cdot \GRR$ in panel c.
    } \label{fig: 5}
\end{figure*}

The example run shown in \cref{fig: 4} converged in 8 steps and was chosen as a representative example. 
To assess the performance of the algorithm more quantitatively, we repeat the tuning procedure 15 times and initialize it at different starting values of the gate voltage $\VABS$. We limit each run to a maximum of 25 measurements (not including the initial measurement) in order to restrict the total run time of the experiment. We chose the number of iterations based on the operation time of the presented experiment. In general, this variable can be set by the user and will be specific to the data collection time and other experimental time scales. The algorithm is run at zero external magnetic field as all measurements shown above.

We calculate the time elapsed between the start and end of each run and plot histograms for different starting $\VABS$ values in \cref{fig: 5}a. 
The dark blue bars indicate runs where the algorithm identified a sweet spot, by finding $f(\VABS) < 0.01$ within 25 measurements. The sea green bars indicate runs where the algorithm identified a sweet spot, with higher tolerance, $0.01<f(\VABS)<0.021$, within 25 measurements. For the light blue bars, the algorithm did not converge with $f(\VABS)<0.021$ within 25 measurements. We note that this does not exclude a sweet spot, as will be discussed below.
Most of the runs that start from $\VABS = \SI{70}{\milli\volt}$ do not converge within 25 measurements. In contrast, nearly all runs starting from $\VABS = \SI{130}{\milli\volt}$ do converge, most of them within 20 minutes. This is a result of limiting each run to 25 measurements, as runs that start from $\VABS = \SI{70}{\milli\volt}$ are farther away from a sweet spot, and need more iterations to converge at a sweet spot. 

In \cref{fig: 5}b, we show histograms of the $\VABS$ values for which the algorithm found the lowest $f(\VABS)$. For starting $\VABS$ values of $\SI{70}{\milli\volt}$, $\SI{80}{\milli\volt}$ and $\SI{90}{\milli\volt}$ (rows I-III), most runs end at $\VABS \approx \SI{90.5}{\milli\volt}$. We can inspect the CSDs measured at these gate values to determine whether they correspond to sweet spots. \cref{fig: 5}c shows $\GLL \cdot \GRR$, for varying $\VLM$ and $\VRM$ at ending $\VABS$ values indicated by the vertical dashed lines in panel b. We see that for \cref{fig: 5}c.I-III, the QD levels hybridize to form a cross, which is consistent with a sweet spot. The corresponding $\GRL$ measurements in \cref{fig: 5}d show both signs of non-local conductance, which confirms that $t=\Delta$~\cite{dvir2023realization}. 
Most of the runs starting from $\VABS = \SI{100}{\milli\volt}$ (row IV) converge in 20 minutes, with the majority ending at $\VABS = \SI{96.5}{\milli\volt}$. From panels c and d, we see that the QD levels do not form a cross, and the non-local conductance is mostly positive. We conclude that the algorithm incorrectly identifies this charge stability diagram as a sweet spot. The confirmation that the algorithm-discovered sweet spot indeed hosts PMMs can be performed via spectroscopic measurements~\cite{dvir2023realization}. We provide the spectroscopic data for both Device A and B in Appendix~\ref{app:validation}.

Runs that start from $\VABS = \SI{110}{\milli\volt}$ (row V) converge with $0.01<f(\VABS)<0.021$ at $\VABS = \SI{99.5}{\milli\volt}$ and with $f(\VABS) < 0.01$ at $\VABS = \SI{123.5}{\milli\volt}$. We can see that the latter is close to a sweet spot, with $t \gtrapprox \Delta$, as seen from the predominantly negative values of $\GRL$. The peak at $\VABS = \SI{99.5}{\milli\volt}$ is close to the $\VABS$ ending values of the runs that started at $\VABS = \SI{100}{\milli\volt}$ runs, and does not correspond to a sweet spot. We define $\SI{90.5}{\milli\volt}$ and $\SI{123.5}{\milli\volt}$ as sweet spots, based on the charge stability measurements. We note that the runs that did not converge with $f(\VABS)$ below tolerance (light blue) are clearly peaked around the sweet spot at $\SI{90.5}{\milli\volt}$. We attribute the higher value of $f(\VABS)$ here to a weaker interaction strength than at the other sweet spot at $\SI{123.5}{\milli\volt}$. Averaging all runs over the varying starting $\VABS$ values, 67.6\% of runs converges in the immediate vicinity of a sweet spot within $\pm \SI{1.5}{\milli\volt}$, and 80.9\% of runs converges within $\pm \SI{4.5}{\milli\volt}$ of an independently verified sweet spot. Averaging over all starting $\VABS$ values, the algorithm converges within $\pm \SI{4.5}{\milli\volt}$ of a sweet spot in 45 minutes. This is sufficiently fast to be beneficial to experimentalists that are tuning Kitaev chain devices into a sweet spot.

 \subsection*{Finite Field Algorithm Performance}
 \begin{figure*}[ht!]
    \centering
    \includegraphics[width=\linewidth]{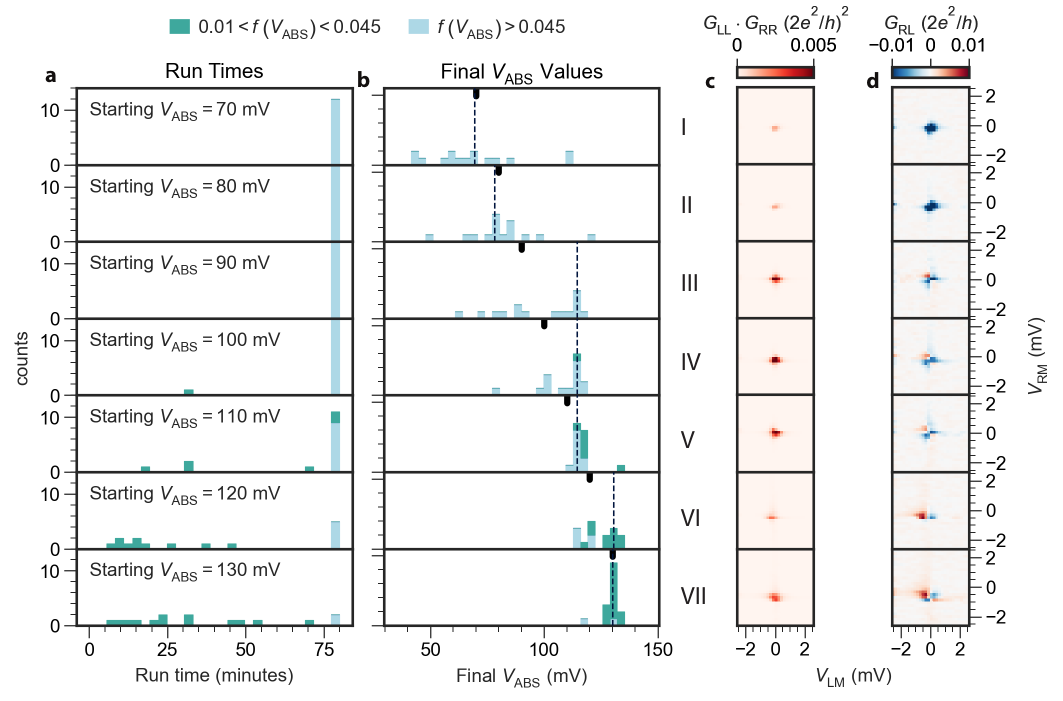}
    \caption{\textbf{Run times and $\VABS$ values at which the algorithm converged for $B=\SI{150}{\milli\tesla}$.} We set a learning rate $\eta = \num{2e-4}$, a momentum $k=0.5$ and a tolerance $\tau = \num{1e-2}$ 
    \textbf{a.} Histograms of the run time of the gradient descent algorithm, for different starting $\VABS$ values. The sea green bars indicate which runs converged with an objective function $0.01<f(\VABS)<0.021$. The light blue bars indicate runs that did not converge with $f(\VABS)<0.021$ within 25 iterations.
    \textbf{b.} Histograms of the $\VABS$ values for which the algorithm found the lowest $f(\VABS)$, for different starting $\VABS$ values. The black marks at the top of the figures indicate the starting $\VABS$ value. The dashed, vertical line indicates the $\VABS$ value closest to the greatest histogram peak with the lowest $f(\VABS)$.
    \textbf{c.} $\GLL \cdot \GRR$, for varying $\VLM$ and $\VRM$, measured at the $\VABS$ value indicated by the dashed, vertical line in panel b.
    \textbf{d.} $\GRL = dI_\mathrm{R}/dV_\mathrm{L}$ for varying $\VLM$ and $\VRM$, for different starting $\VABS$ values. The data was measured simultaneously with $\GLL \cdot \GRR$ in panel c.
    }\label{fig: 6}
\end{figure*}

In the previous section, we found sweet spots in the absence of an external magnetic field. In this case, the emerging Poor Man's Majorana zero modes are Kramers pairs, which retain the protection against local perturbations, 
but have an additional degeneracy due to time-reversal symmetry~\cite{bozkurt2024}. This degeneracy has no impact on the demonstration of the automated tuning algorithm in the previous section, as the CSDs look identical to the spin-polarized result when the charging energy of the QDs is large compared to the parent gap. We now break time-reversal symmetry by applying an external field $B=\SI{150}{\milli\tesla}$ along the nanowire axis, to test whether our algorithm works both in the absence and presence of magnetic field. The magnetic field Zeeman-splits the ABSs in addition to the QDs and lowers their energy. To compensate for the resulting increased hybridization between the QDs and ABS, we raise the tunnel barriers between them using electrostatic gates.

We repeat the automated tuning algorithm for 15 times at different starting values of voltage $\VABS$ each. We, again, limit each run to a maximum of 25 measurements (not including the initial measurement), and show histograms of the run times in \cref{fig: 6}a. In \cref{fig: 6}a.I-III, we see that none of the runs converge with $f(\VABS)<0.045$ within 80 minutes. The corresponding histograms in \cref{fig: 6}b show a broader distribution than for $B=\SI{0}{\tesla}$, as seen in \cref{fig: 5}b. The runs that start from $\VABS$ values of \SI{110}{\milli\volt}, \SI{120}{\milli\volt} and \SI{130}{\milli\volt} (rows V-VII) converge with $0.01<f(\VABS)<0.045$ more frequently, and the resulting histograms in \cref{fig: 6}b are more clustered. 

 The CSDs in \cref{fig: 6}c,d, indicate that the interaction between the QDs is significantly weaker than for $B=\SI{0}{\tesla}$, as seen from the smaller avoided crossings. We attribute this to the stronger tunnel barriers we had to set to compensate for the lower ABS energy. This leads to generally larger values of $(t-\Delta)/(t+\Delta)$, as the denominator becomes smaller. In addition, our regularization procedure of the CNN predictions leads to a larger $f(\VABS)$ when the conductance is low. 
 
 We attribute the worse convergence of the algorithm to the weaker QD-QD interaction. Nonetheless, we see from \cref{fig: 6}d that a number of peaks of the histograms in panel b correspond to sweet spots. We further note that $\GLL \GRR$ is the input for the CNN, which does not have the added information of $\GRL$. Based on the CSDs measured using $\GLL \GRR$ in \cref{fig: 6}c alone, it is hard for an experimentalist to identify a sweet spot. Defining $\SI{114.26}{\milli\volt}$ and $\SI{130.44}{\milli\volt}$ as sweet spots based on the measurements of $\GRL$, we find that 45.7\% of the runs converge within $\pm \SI{1.62}{\milli\volt}$, and 60\% converge within $\pm \SI{4.85}{\milli\volt}$ of these sweet spots. In this case, the algorithm can be run numerous times to gather statistics. The value of $\VABS$ that is converged upon most frequently will then indicate an optimal sweet spot. While the performance at finite magnetic field is worse than the algorithm's performance at $B=\SI{0}{\tesla}$, we note that the CNN was trained only on strongly interacting data (see \cref{app:training}). We also note that we chose the same tolerance for $f(\VABS)$ as for \cref{fig: 5}, while the QD-QD interaction was weaker. These two aspects mean that the algorithm will find less $f(\VABS)$ values below tolerance than before. We conclude that the algorithm can partially generalize to previously unseen experimental data in the weakly interacting regime. It can then be run multiple times to statistically identify sweet spots.

\subsection*{Finite Field CNN Validation}
\begin{figure}[ht!]
    \centering
    \includegraphics[width=\linewidth]{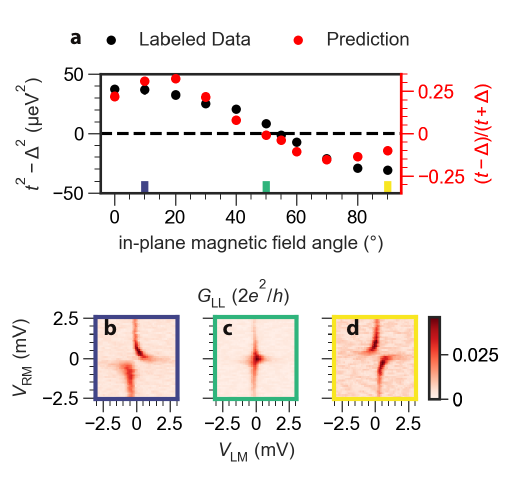}
    \caption{\textbf{Convolutional neural network (CNN) predictions of charge stability diagrams (CSDs) measured at $B=\SI{80}{\milli\tesla}$ on the device investigated in the main text of~\cite{ten2024two}} 
    \textbf{a.} Labeled and predicted couplings for varying in-plane magnetic field angle at fixed $\VABS$. Black markers: $t^2-\Delta^2$ obtained by calculating the distance between the hyperbolea in measured CSDs, adapted from Fig.~2d of~\cite{ten2024two}. Red markers: $(t-\Delta)/(t+\Delta)$ predictions by the CNN of the CSDs, after training on the device B data.    
    \textbf{b-d.} $\GLL$, for varying $\VLM$ and $\VRM$, measured at values of the in-plane magnetic field angle indicated by the colored ticks in panel a.
    }\label{fig: 7}
\end{figure}

To further evaluate the CNN classifier’s performance on finite field data, we test it on measurements from an additional device. Specifically, we use data collected at finite magnetic field from a second 2DEG device—referred to as device B in Ref.~\cite{ten2024two}—which is identical in design to the device B used in this work. \cref{fig: 7}a displays the values of $t^2 - \Delta^2$ extracted from charge stability diagrams (CSDs) measured at $B = \SI{80}{\milli\tesla}$.
As bias-spectroscopy was not performed for each in-plane magnetic field angle, we cannot independently extract $t$ and $\Delta$. However, we can still extract and compare the quantity $t^2 - \Delta^2$. The point $t^2-\Delta^2=0$ corresponds to a sweet spot, as can be seen from the corresponding CSD in \cref{fig: 7}c, combined with the avoided crossings pointing along different diagonals in \cref{fig: 7}b,d. 

The red markers in \cref{fig: 7}a show the values of $(t-\Delta)/(t+\Delta)$ predicted by the CNN based on the measured CSDs. We see that the CNN predictions closely follow $t^2-\Delta^2$, albeit with a difference in scale. The equality 
$$t^2-\Delta^2= (t+\Delta)^2 \cdot \frac{t-\Delta}{t+\Delta}$$
implies that the two quantities differ by a factor of $(t+\Delta)^2$. 
As there is strong agreement between the labeled data and predictions (black and red dots in \cref{fig: 7}a) and the transition point is predicted correctly, we conclude that the CNN does work at finite magnetic fields. 
We note that the CSDs in \cref{fig: 7}b-d show stronger interaction than the corresponding ones of \cref{fig: 6}c,d. 
The worse performance of the tuning algorithm at finite field can possibly be explained by a lack of weakly-interacting quantum dot data in the training sets.




\section*{Discussion}
Currently, the algorithm converges within $\pm \SI{4.5}{\milli\volt}$ of a sweet spot in 45 minutes, which is generally faster than the time duration for manually tuning to a sweet spot. Since the CNN predictions and gradient descent algorithm calculations are nearly instantaneous, the run time is dominated by the DC conductance measurements. Performing RF reflectometry measurements is significantly faster and was used to explore a multidimensional parameter space in semiconductor-superconductor devices before~\cite{wang2022parametric}. We believe that the algorithm can converge significantly faster, provided that it is possible to train the CNN on RF reflectometry data generated from theoretical models instead of DC conductance presented here. Also, we note that we start the algorithm from tuned-up QDs in a region of $\VABS$ where ABSs are visible in the spectroscopy measurements. Ideally, the algorithm should also include tuning the QDs and ABSs to have interaction, which also takes a significant amount of time. We note that there are several automated tuning algorithms for forming double quantum dots in literature~\cite{darulova2020autonomous, moon2020machine, zwolak2020autotuning,kalantre2019machine}. Adapting these algorithms and integrating them into the tuning routine presented here would additionally increase speed of Kitaev chain devices tune-up.

\section*{Outlook}
We have demonstrated the automatic tune-up of a minimal, two-site Kitaev chain. 
While this is a necessary first step, longer chains are needed to realize a topological phase transition~\cite{sau2012realizing} or a parity qubit with appreciable coherence times~\cite{leijnse2012parity,Pan2025}.
Extending the chain to host Poor Man's Majorana zero modes at its edges requires tuning additional couplings between QDs. 
Crucially, the Kitaev chain model only demands that the condition $t=\Delta$ be satisfied pairwise, between adjacent sites~\cite{fulga2013adaptive, kitaev2001unpaired}. 
This reduces the tuning of an $N$-site Kitaev chain to setting $N-1$ pairwise couplings to $t=\Delta$. 
Since our algorithm can successfully tune a two-QD system to $t=\Delta$, it can be extended to Kitaev chains of arbitrary lengths under the right conditions.

\begin{figure*}[ht!]
    \centering
    \includegraphics[width=\linewidth]{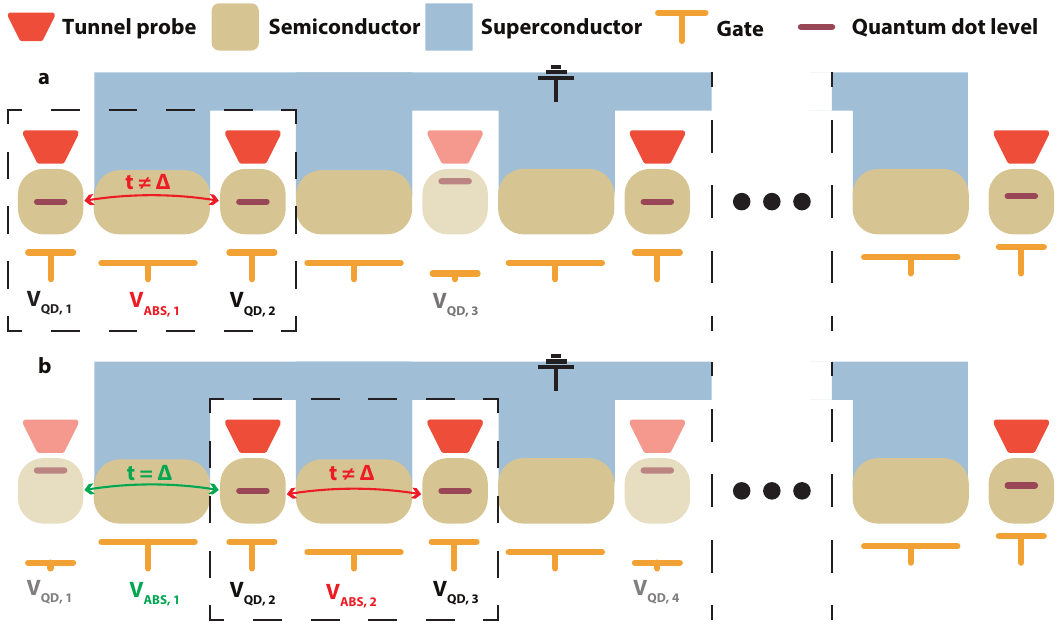}
    \caption{\textbf{Protocol for automated tuning of Kitaev chains of arbitrary length, based on sequentially tuning unit cells to $t=\Delta$.} Orange-red trapezoids indicate tunnel probes, beige rectangles indicate semiconductor sections, blue is the grounded superconductor, yellow electrostatic gates and brown QD levels. Tunnel barriers between the QDs and hybrid sections are omitted for clarity. 
    \textbf{a.} Step 1 in the tuning algorithm, in which none of the pairwise interactions is sweet spot ($t \neq \Delta$). The leftmost two QDs are tuned into the sweet spot by the algorithm, by varying $\VABSi{1}$. The tuning unit cell is indicated by the dashed rectangle. This pair of QDs is isolated from the rest of the chain by detuning the next QD above the gap using $\mathrm{V}_{\mathrm{QD}, 3}$.
    \textbf{b.} Step 2 in the tuning algorithm, where the first pair now fulfills $t=\Delta$. The tuning unit cell is now isolated from other parts of the chains by detuning both QD 1 and 4. 
    }\label{fig: 8}
\end{figure*}

In \cref{fig: 8}, we propose how our pairwise sweet spot tuning algorithm can be extended to become an automated N-site Kitaev chain tuning protocol. This builds on the assumption that tuning one unit cell to the $t=\Delta$ condition is similar to tuning a minimal Kitaev chain to the sweet spot, which has been proposed in Ref.~\cite{liu2025scaling}.
The proposed device is a 2DEG with normal leads acting as tunnel probes for each QD. 
The hybrid sections between the QDs host ABSs and are proximitized by a single, grounded superconductor. 
In addition to tunnel barriers, local electrostatic plunger gates can be used to control the hybrid section and QD electrostatic potentials. 


In \cref{fig: 8}a, starting from the left end of the chain, CSDs of the first QD pair can be measured using the corresponding tunnel probes. 
For large enough detuning of the third QD, the leftmost pair reduces to an effective two-site Kitaev chain~\cite{bordin2025enhanced}. 
The algorithm presented in this work can then be used to set the first pair to $t=\Delta$ within this tuning unit cell indicated by the dashed rectangle. 
In finding a $\VABSi{1}$ value corresponding to a sweet spot, we effectively choose a phase for the couplings that needs to be matched by every subsequent pair in the chain.

In \cref{fig: 8}b, the leftmost pair of QDs has been tuned to the sweet spot. 
Before tuning the interaction between QDs 2 and 3, we isolate the pair by detuning QDs 1 and 4. 
The tuning unit cell has been moved forward by one step, and can be treated as an effective two-site Kitaev chain again. 
The algorithm can then be used to find a $\VABSi{2}$ value for which $t=\Delta$ for this pair. In the example of \cref{fig: 8}b, this protocol is implemented as follows. First,
$\VABSi{2}$ is varied until a sweet spot between QD2 and QD3 is reached. 
Then, $\mathrm{V}_\mathrm{QD, 1}$ is set such that the QD level is on resonance with QD2 and QD3, creating a three-site chain. As we have now added a site, the phase difference across the superconducting segments becomes relevant, as it affects the relative phase of tunnel couplings between QD2 and QD3. 
The protection of the Kitaev chain is maximal when there is no phase difference between neighboring sites~\cite{ten2025observation}. 
To probe this phase difference, we follow the steps outlined in~\cite{Liu2025}.
First, $\mathrm{V}_\mathrm{QD, 3}$ is varied while the differential conductance of QD3 is measured. 
In the case of a robust zero-bias conductance peak, a $0$-phase has formed, and the next pair (QD3-QD4) can be added to the chain.
In the case of a zero-bias peak in the local conductance of QD3 that splits due to $\mathrm{V}_{\mathrm{QD, 3}}$, a sweet spot at a different $\mathrm{V}_{\mathrm{ABS, 2}}$ value is needed to obtain a relative $0$-phase.

When no magnetic flux is threaded through the loop between the first two hybrid sections, a phase difference can still arise between the $t$ and $\Delta$ terms of the two QD pairs~\cite{sau2012realizing} (either 0 or $\pi$).
In the case of a $\pi$-phase difference, a domain wall is created, preventing the formation of a gapped bulk in the chain~\cite{bordin2025enhanced, ten2025observation}. As a result, the MZMs lose their protection against local perturbations given by longer Kitaev chains.
Having tuned to a $t=\Delta$ sweet-spot for a newly added QD, we must thus check the relative phase of the newly found sweet-spot.
This can be achieved through finite bias spectroscopy, as detailed in~\cite{liu2025scaling,huisman2026}.
If the phase constitutes a $\pi$-phase, all that is needed to remove the domain wall is to repeat the tuning protocol for a sweet-spot at a neighboring $\VABS$ value, which induces another $\pi$-shift.

When using this additional protocol step, a sign-ordered Kitaev chain of arbitrary size can be tuned to host MZMs at its edges without an external flux~\cite{liu2025scaling}. 
This can be combined with the algorithm presented in this work to do so autonomously.

\section*{Conclusion}
We have shown that a supervised machine learning algorithm can predict $(t-\Delta)/(t+\Delta)$ from the charge stability diagrams of a minimal Kitaev chain. The algorithm was first trained on data generated from a theoretical model, and was shown to generalize well to conductance measurements of a two-dimensional electron gas device. After retraining on a part of this experimental data, the predictions of $(t-\Delta)/(t+\Delta)$ became significantly closer to the experimentally extracted values. The retrained algorithm demonstrated strong generalization to conductance measurements obtained from a nanowire device, achieving good accuracy across different architectures.

A gradient descent algorithm was then used to drive the nanowire device into a sweet spot by changing $\VABS$, based on the predictions of $(t-\Delta)/(t+\Delta)$. Note that the CNN was not retrained on data from this device. When the QD-QD interaction is strong, the algorithm converges at two sweet spots with different $t+\Delta$ values, which is reflected in the value of $f(\VABS)$ at convergence. At finite external magnetic field, we had to decrease the QD-QD interaction because of the lower ABS energy. As a result fewer runs converged, which we attribute to the bias towards strong QD-QD interaction in the training and retraining data. However, a number of peaks in the ending $\VABS$ histogram coincide with sweet spots, which shows that the CNN does, to a degree, generalize to weak interaction. When benchmarking the CNN on a previously-unseen finite field dataset from two-dimensional electron gas device, its predictions matched well with the experimentally-obtained data. From this, we conclude that the CNN can fundamentally work at finite field, as long as the QD-QD coupling is strong enough.

Finally, we present a protocol to autonomously tune up Kitaev chains of arbitrary lengths automating the protocol proposed in Ref.~\cite{liu2025scaling}.
Our automated tuning algorithm can successfully drive a minimal Kitaev chain into a sweet spot. Tuning and monitoring multiple sweet spots is required for quantum information experiments involving Majorana zero modes. Our work paves the way for tuning up more complicated Kitaev chain devices that are required for braiding, anyon fusion and other quantum information experiments \cite{liu2023fusion,tsintzis2024majorana}.

\FloatBarrier

\section*{Acknowledgements}
This work has been supported by the Dutch Organization for Scientific Research (NWO), Microsoft Corporation Station Q, Academy of Finland Projects No. 331342 and No. 358088 and the Jane and Aatos Erkko Foundation. This publication is part of the project Engineered Topological Quantum Networks (with Project No. VI.Veni.212.278) of the research program NWO Talent Programme Veni Science domain 2021 which is financed by the Dutch Research Council (NWO). S.G. and M.W acknowledge financial support from the Horizon Europe Framework Program of the European Commission through the European Innovation Council Pathfinder grant no. 101115315 (QuKiT). We thank Di Xiao, Candice Thomas and Michael J Manfra for providing the 2DEG wafers.

\section*{Author contributions}
DvD, FZ, AB, GPM, NvL, GM, BR, GW and JCW fabricated the nanowire device. 
SLDtH, IK and QW fabricated the two-dimensional electron gas device.
VS, SLDtH and DvD performed the electrical measurements. 
RK and JLL designed and trained the neural network, with input from EG.
SLDtH, RK, DvD and EG designed the experiment. 
SLDtH, RK, DvD analyzed the data. 
DvD, SLDtH, RK, JLL and EG prepared the manuscript with input from all authors. 
TD, S.~Goswami, JLL, LPK and EG supervised the project. 
CXL and AMB developed the theoretical model with input from MW. 
GB, S.~Gazibegovic, and EPAMB performed InSb nanowire growth.
\section*{Data availability}
All raw data in the publication, analysis code, plotting code, as well as spectroscopy labelling code is available at \url{https://doi.org/10.5281/zenodo.10900882}.
This paper is supplemented by a GitLab repository with all the code and data necessary to reproduce our results available at \url{https://gitlab.com/QMAI/papers/crossplatformkitaev}.
\appendix
\renewcommand{\thesubsection}{\Roman{subsection}} 
\setcounter{figure}{0}
\renewcommand\thefigure{S\arabic{figure}}

\section{CNN ARCHITECTURE}\label{app:architecture} 

Convolutional Neural Networks (CNNs) are designed for efficient data analysis, especially for visual tasks such as image recognition. Unlike traditional fully-connected neural networks (NNs), CNNs utilize convolutions to process information in image-like datasets. These enhance their capacity to handle spatially-correlated information present among pixels in image-like datasets~\cite{Li2022,oshea2015introduction}. 
The architectural framework of CNNs leverages spatial relations within the data through locally-connected layers that improve the computational efficiency by neglecting correlations between distant data points.

At the foundation of CNNs are trainable convolutional filters that are crucial for capturing spatial correlations in the data set. The filter size is a hyper parameter of the network architecture and, therefore, adaptable to specific problem sets. Furthermore, to reduce the complexity and number of parameters, CNNs integrate dimensionality reduction techniques such as pooling operations~\cite{Murray_2014_CVPR}. These operations allow us to preserve essential features and increase computational efficiency at the same time.

The architecture of the CNN to make prediction in this work is shown in \cref{tab:CNN_architecture}
The network architecture employed in this study comprises a total of 9 layers. The initial 6 layers constitute the convolutional segment of the network, followed by a flattening operation and the subsequent utilization of a fully-connected NN.
The initial 6 layers operate within a dual-input framework, where two independent and unconnected CNNs can process different images, such as $G_{LL}$ and $G_{RR}$ (or non-local conductance $G_{LR/RL}$), and integrate them in the subsequent NN segment. Notably, in the current research, the dual-input framework is not exploited, as we pre-process the data by multiplying $G_{LL}$ with $G_{RR}$. 
However, the dual-input framework is kept for the sake of generalization that allows us to adapt the CNNs and tuning algorithm to different inputs in future research.


\begin{table}[h]
    \centering
    \caption{Architecture of the CNN.\\
    {\footnotesize *we are using a dual-input framework for the first 6 layers}}
    \label{tab:CNN_architecture}
    \begin{tabularx}{.7\columnwidth}{|X|c|}
        \hline
        \multicolumn{2}{|c|}{CNN Classifier} \\
        \hline
        \hline
        Layer & Output shape \\
        \hline
        Input &  (28,28,1) \\
        Conv2D &   (28, 28, 32) \\
        MaxPooling2D &  (14, 14, 32) \\
        Conv2D &   (14, 14, 64) \\
        MaxPooling2D & (7, 7, 64) \\
        Flatten &  (3136) \\
        \hline
        Concatenate $^*$ & (6272) \\
        Dense & (128) \\
        Dense & (1)\\
        \hline
        total parameters & 840,705 \\
        \hline
    \end{tabularx}
\end{table}

\section{CNN TRAINING}\label{app:training} 

In this work, we are utilizing a two-step training process for the CNN. The first step involves training with a dataset generated through numerical calculations, while the second step involves re-training using a smaller set of experimental conductance data. 

\subsection*{Kitaev Chain Effective Model}
We are modeling the double quantum dot system coupled by a semiconductor-superconductor hybrid segment as a two-site Kitaev chain. The couplings between the quantum dots are mediated by ABSs in the hybrid segment. 
From a theory point of view, the system can be seen as 3-site model where the hybrid segment gets treated as site. We are following the work of Ref~\cite{Liu2024} and refer there for a more detailed description of the theory model.
The corresponding Hamiltonian of the 3-site model takes the form~\cite{liu2022tunable,Tsintzis2022,Domnguez2016} 
\begin{equation}
    \begin{split}        
    H =& \hspace{1.5mm} H_D + H_S + H_T \\[1mm] 
    H_D =& \sum_{a=L,R} \left( \varepsilon_{Da} + E_{ZDa} \right) n_{Da\uparrow} + \left( \varepsilon_{Da} - E_{ZDa} \right) n_{Da\downarrow}\\[-1mm] 
    & \hspace{10.5mm}+ U_{Da} n_{Da\uparrow} n_{Da\downarrow}  \\[1.5mm]
    H_S =& \hspace{1.5mm} \varepsilon_A (n_{A\uparrow} + n_{A\downarrow}) + \Delta_0 (c_{A\uparrow} c_{A\downarrow} + c_{A\downarrow}^{\dagger} c_{A \uparrow}^{\dagger})  \\[1.5mm]
    H_T =& \sum_{\sigma=\uparrow,\downarrow} ( t_L c^{\dagger}_{A\sigma} c_{DL\sigma} + \sigma t_{Lso} c^{\dagger}_{A\sigma} c_{DL\sigma} \\[-1.5mm]
     & \hspace{8mm} + t_R c_{DR\sigma}^\dagger c_{A\sigma} + \sigma t_{Rso} c^\dagger_{DR\sigma} c_{A\sigma} ) + h.c. \, .
    \end{split} \label{eq:fullH}
\end{equation}
The Hamiltonian can be split into three parts, the quantum dot Hamiltonian, $H_D$, the hybrid segment with the ABSs, $H_S$, and the tunnel coupling between ABS and quantum dots, $H_T$. 
$H_D$ contains the orbital energy, $\varepsilon_{Da}$, the electron occupation, $n_{Da\sigma}$, the Zeeman energy, $ E_{ZDa}$, and Coulomb repulsion, $U_{Da}$. The index $a$ defines the left/right quantum dot.
The hybrid Hamiltonian $H_S$ contains the normal state energy $\varepsilon_A$ and an induced pairing gap $\Delta_0$.
The tunnel Hamiltonian $H_T$ includes two couplings between the quantum dot and ABS: a spin-conserving, $t$, and a spin-flipping process, $t_{so}$. The spin-orbit interaction determines the ratio of $t_{so}/t$.

Proximity effects of the semiconductor-superconductor hybrid transform the quantum dot orbitals into Yu-Shiba-Rusinov states (YSR)~\cite{YULUH1965,Shiba1968,rusinov1969theory} and create a new basis of spinless fermions for a Kitaev chain model. The YSR states are a superposition of electron and hole components and in this basis, the Kitaev chain has more generalized effective couplings, describing the interaction between the two YSR states. 
We can write the effective Hamiltonian describing the interaction between these two states as
\begin{equation}
    H_{\text{eff}}^{\text{coupling}} = \sum_{\sigma,\eta=\uparrow,\downarrow} \left( t_{\sigma\eta} c_{DL\sigma}^{\dagger} c_{DR\eta} + \Delta_{\sigma\eta} c_{DL\sigma}^{\dagger} c_{D R \eta}^{\dagger} \right) + h.c. \, ,
\end{equation}
where we consider different ECT and CAR amplitudes, $t_{\sigma\eta}$ and $\Delta_{\sigma\eta}$ between electron and hole components of the quantum dots. Note that each site still as an on-site energy and Coulomb interaction as in Eq.~\eqref{eq:fullH}.
Considering that the YSR states have electron and hole components, the couplings $t_{\sigma\eta}$ and $\Delta_{\sigma\eta}$ have to be generalized. 
The ground states of a single proximitized quantum dot are a spin singlet and a spin-down state in the even- and odd-parity subspace:
\begin{equation}
    \begin{split}
        |S\rangle &= u|00\rangle - v|11\rangle, \quad |\downarrow\rangle = |01\rangle \\[1mm]
        u^2 & = 1 - v^2 = \frac{1}{2} + \frac{\xi}{2E_0} \\
        \xi &= \varepsilon + \frac{U}{2}, \quad E_0 = \sqrt{\xi^2 + \Delta_{\text{ind}}^2} \, .
    \end{split}
\end{equation}
This allows us to define the YSR state as $|\downarrow\rangle=f_{YSR}^\dagger |S\rangle$ and the effective coupling of Eq.(B2) written in terms of YSR states becomes
\begin{equation}
    H_{\text{eff}} = \sum_{a=L,R} \delta \varepsilon_a f_a^\dagger f_a + \Gamma_o f_L^\dagger f_R + \Gamma_e f_L^\dagger f_R^\dagger + h.c. \, ,
\end{equation}
where $\delta \varepsilon_a=E_\downarrow-E_S$ is the excitation energy of the YRS state and $\Gamma_{o/e}$ are odd/even generalized effective couplings between the YSR states. The odd coupling is defined as
\begin{equation}
    \begin{split}
        \Gamma_o &= \langle S \downarrow | H_{\text{eff}}^{\text{coupling}} | \downarrow S \rangle \\
        &= -t_{\uparrow \uparrow} v_L v_R + t_{\downarrow \downarrow} u_L u_R + \Delta_{\uparrow \downarrow} v_L u_R - \Delta_{\downarrow \uparrow} u_L v_R \, ,     
    \end{split}
\end{equation}
where $|\downarrow S\rangle$ and $|S\downarrow\rangle$ are odd parity tensor states and $\Gamma_o$ is a linear combination of spin-conserving couplings, namely equal spin ECT and opposite spin CAR processes.
The even-parity coupling
\begin{equation}
    \begin{split}
        \Gamma_e &= \langle SS | H_{\text{eff}}^{\text{coupling}} | \downarrow \downarrow \rangle \\
        &= -\Delta_{\uparrow \uparrow} v_L v_R + \Delta_{\downarrow \downarrow} u_L u_R + t_{\uparrow \downarrow} v_L u_R - t_{\downarrow \uparrow} u_L v_R     
    \end{split}
\end{equation}
couples states with total spin zero and one, i.e. breaks spin conservation.
These odd- and even-parity couplings can be seen as more generalized effective $t$ and $\Delta$ parameter from the Poor Man's Majorana Kitaev chain model~\cite{leijnse2012parity}.
To obtain the differential conductance, we employ the Mahaux–Weidenmuller formula to compute the scattering matrix~\cite{Christiansen2009}. For this, we rewrite the derived effective Hamiltonian of Eq.(B4) in the Bogoliubov-de Gennes formalism, solve the corresponding eigenequation, and obtain the $W$-, scattering matrix and conductance matrix. In this work, we focus on the diagonal elements of the conductance matrix $G_{LL}$ and $G_{RR}$.

\subsection*{Numerical Training Data}

\begin{figure}[ht!]
    \centering
    \includegraphics[width=\linewidth]{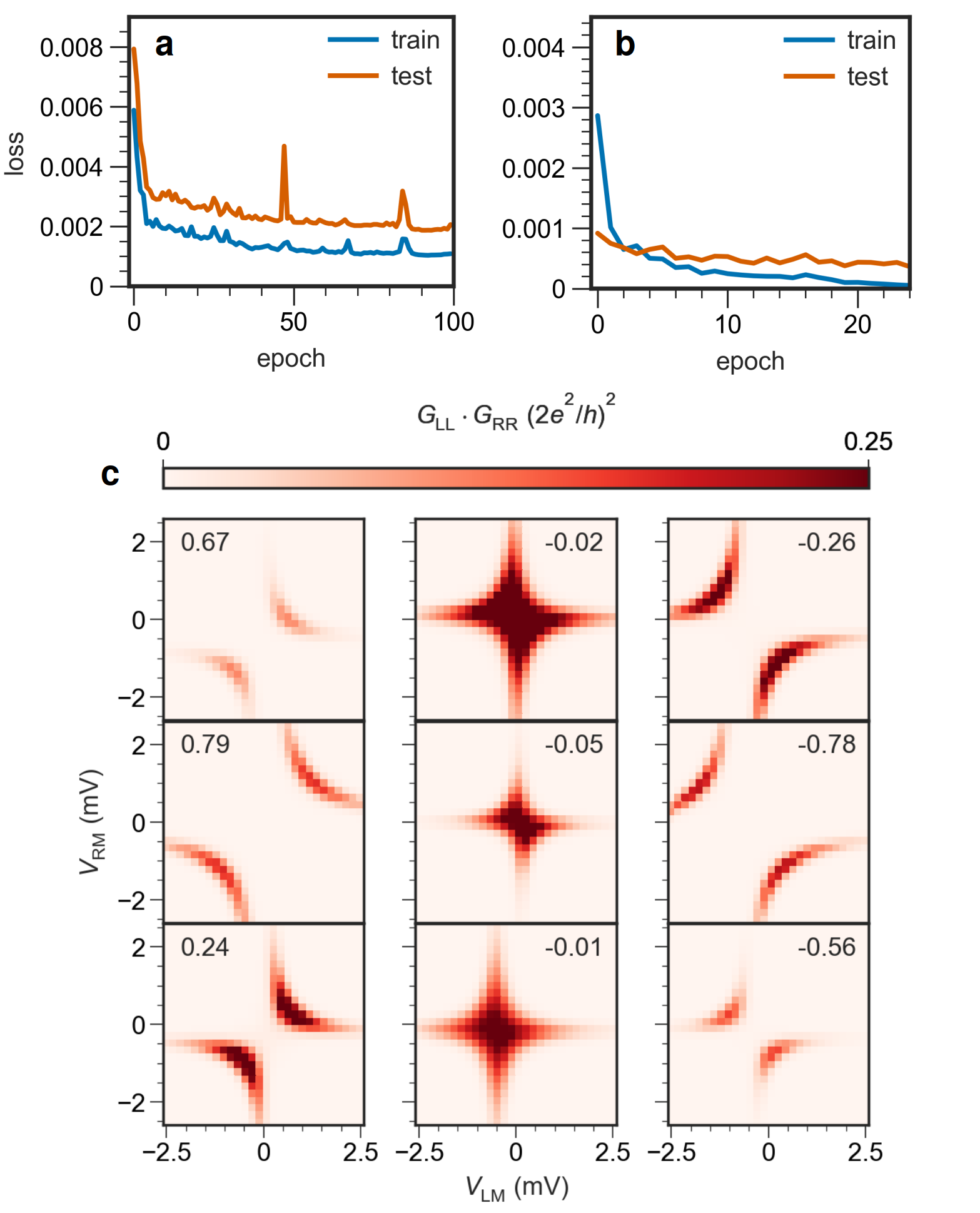}
    \caption{\textbf{Training evaluation and numerical training data of the convolutional neural network (CNN)} 
    \textbf{a.} The train- and test-loss during the initial training of the CNN with the numerical data set (shown in c). 
    \textbf{b.} The train- and test-loss during the re-training of the CNN with labeled experimental data (shown in Fig.~\ref{fig: 3}(a-e)
    \textbf{c.} Numerical calculations of the product of the left and right differential conductance, $\GLL \cdot \GRR$, for varying $(t-\Delta)/(t+\Delta)$, including the ECT-dominates, sweet spot, and CAR-dominated regime.
    } \label{fig: training}
\end{figure}

After introducing the theory model, we create the theoretical training data by generating a diverse set of Hamiltonians of Eq.(B1) leading to an effective Hamiltonian in the YSR basis of Eq.(B4), incorporating random variations within of parameters in the model in specified intervals:
\begin{itemize}
    \item induced Zeeman energy in each quantum dot $ E_{ZDa}\in I=[-0.2,0.2]$
    \item induced pairing gap $\Delta_0 \in I=[0.2,0.6]$
    \item tunneling couplings $t_{L/R} \in I=[0.05, 0.4]$
    \item $t's$ and $\Delta's \in I=[0.01,0.48]\\[1mm] \rightarrow \Gamma_{odd/even} \in I=[0.01, 0.48]$
    \item temperature $T \in I=[0.02,0.03]$ 
    \item left/right lead coupling $\Gamma_{L/R} \in I=[0.04, 0.06]$
\end{itemize}
While we include a Zeeman energy on the QDs, we do not consider the up state of the QDs. As a result, the Zeeman energy only lowers the energy of the down states. 
The theoretical data, depicted in Fig.~\ref{fig: training}(c), closely resembles the experimental conductance data including, e.g., variations in lead couplings, background noise, and interaction strengths across a broad parameter range spanning from the ECT- to the CAR-dominated region. 
We illustrate in Fig.~\ref{fig: training}(a) the training/test loss for the initial training step. The loss function shows the expected behavior, with the test-loss consistently higher than the training-loss, reaching saturation at around 75 epochs. We have chosen a training set of 6000 samples and trained the CNN with a batch size of 16 for the 100 epochs. The fluctuations during the training can be related to the stochastic nature of the gradient descent algorithm~\cite{Bottou2010}, the small batch size, and randomly-chosen training/test data.

\subsection*{Experimental Training Data}
We show the training/test loss for the re-training with an experimental training set in Fig.~\ref{fig: training}(b). The test-loss is again consistently higher than the training-loss and saturates at around 20 epochs. For the re-training, we only train the CNN for 25 epochs, starting from the already-trained CNN parameters and fine-tuning the parameters to experimental data.

In Figs.~\ref{fig: 3}(a,b), we compare the CNN predictions for the experimental dataset. In (a), we show the predictions of the initial theory-only CNN. The predictions show a constant gradient for increasing $\VABS$ capturing the experimental values overall with good accuracy.
However, the theory CNN does not predict the sweet spot region around $\VABS = \SI{-622}{\milli\volt}$ well which possibly can lead to convergence problems when approaching the sweet spot with the gradient descent algorithm.
Figure~\ref{fig: 3}(b) shows the prediction of the re-trained CNN for the experimental test-data. In this case, the CNN captures the sweet spot region well. The gradient towards the sweet spot coincides well with the labeled data which is one of the most important factors for CNN-tuning algorithm.
In Fig.~\ref{fig: 3}(c-e) we show three conductance measurements for the ECT, CAR, and sweet spot regime taken at specific $\VABS$ from above.


\begin{figure*}[t!]
    \centering
    \includegraphics[width=\linewidth]{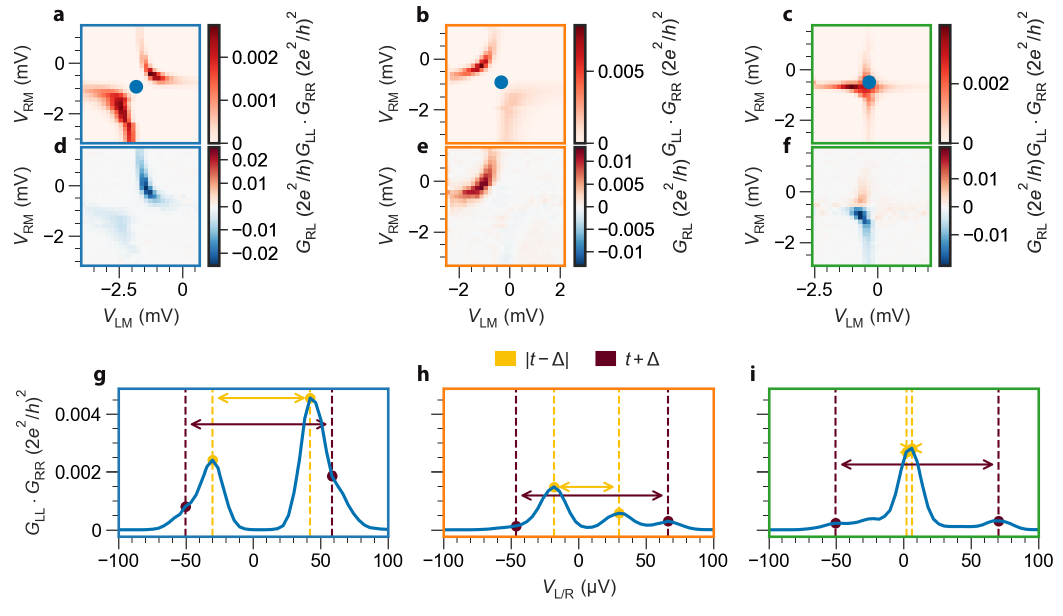}
    \caption{\textbf{Extraction of labels using tunnel spectroscopy.} \textbf{a.-c.} The product of the left and right differential conductance, $\GLL \cdot \GRR$, for varying $\VLM$ and $\VRM$. The measurements were performed at $\VABS$ values of \SI{85}{\milli\volt}, \SI{104}{\milli\volt} and \SI{115}{\milli\volt} respectively.
    \textbf{e.-g.} Non-local conductance, $\GRL = dI_\mathrm{R}/dV_\mathrm{L}$ for varying $\VLM$ and $\VRM$. The data was measured simultaneously with panels b-d.
    \textbf{g.-i.} $\GLL \cdot \GRR$ for varying left/right lead bias, $V_\mathrm{L/R}$, measured at the $\VLM$ and $\VRM$ values indicated by the blue marker in panels a-c. The purple, dashed lines indicate the outer peaks. The yellow, dashed lines indicate the inner peaks.   
    }\label{fig: label procedure}
\end{figure*}
\clearpage

\section{DATA LABELING PROCEDURE}\label{app:label} 
\cref{fig: label procedure}a,c show the charge stability diagrams as presented in \cref{fig: 3}a-c. We use python code to interpolate the center of the avoided crossings, which is indicated by the blue markers. At the $\VLM$ and $\VRM$ values corresponding to these markers, we measure $\GLL$ for varying $\VL$, and $\GRR$ for varying $\VR$, and show the resulting $\GLL \cdot \GRR$ in \cref{fig: label procedure}g-i. Each spectrum consists of two inner peaks separated by $\lvert t-\Delta \rvert$ and two outer peaks separated by $t+\Delta$ (see methods section of~\cite{zatelli2024robust} for details on the model). These are indicated by the yellow and purple arrows in panels g-i. We label the $V_\mathrm{L/R}$ values corresponding to the peaks manually, since $\GLL \cdot \GRR$ can be too low for a peak-finding algorithm to work. Because the inner peaks do not provide information on the sign of $ t-\Delta$, we extract it from the non-local conductance. When $t>\Delta$, elastic co-tunneling dominates over crossed Andreev reflection, and the sign of non-local conductance will be negative \cite{dvir2023realization}. Likewise, $\GRL$ will be predominantly positive for $t<\Delta$, so we extract the sign from:
\begin{equation*}
    sgn(t-\Delta) = \begin{cases}
  1  & -min(\GRL) > max(\GRL)  \\
  -1 & max(\GRL) > -min(\GRL) 
\end{cases}
\end{equation*}
We combine the sign found from the non-local conductance measurements with the inner and outer peak spacing to calculate $(t-\Delta)/(t+\Delta)$ for each $\VABS$ value in \cref{fig: 3}g.

We note that directly relating the peak spacings to the CAR and ECT rates is only justified in the infinite Zeeman limit. At zero magnetic field, in limit of large charging energy on the QDs, the sweet spot condition becomes $t=\sqrt{2}\Delta$ as a result of the larger Hamiltonian basis that now includes both spin species~\cite{bozkurt2024}. At this sweet spot, there is still a peak at the Fermi level and two peaks that indicate the gap. While the distance between the outer peaks is then no longer $t+\Delta$, we continue to use this form of labeling for the CNN. The direct relation with the CAR and ECT rates is then severed. This means that we are not training the CNN on transport rates, but rather on peak spacings. For our purposes this is not problematic, as minimizing $(t-\Delta)/(t+\Delta)$ will still correspond to finding the sweet spot. We note that there can be additional excited states in spectroscopy at zero field~\cite{ten2024two}. These are only visible at one bias polarity. As we look for particle-hole symmetric peaks in $\GLL \cdot \GRR$, we filter out these low-Zeeman excited states.

\section{CNN PREDICTION LIMITATIONS}\label{app:restraints} 
\begin{figure}[ht!]
    \centering
    \includegraphics[width=\linewidth]{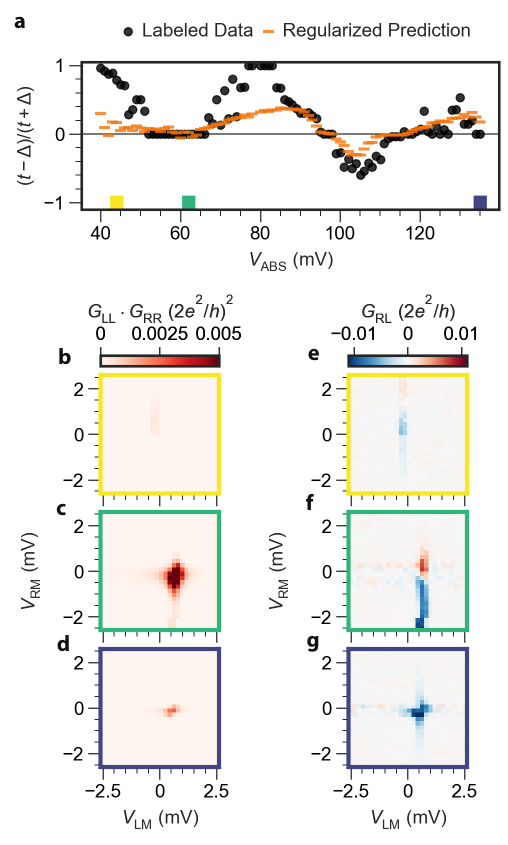}
    \caption{\textbf{Selected charge stability diagrams.} 
    \textbf{a.} $(t-\Delta)/(t+\Delta)$ for varying $\VABS$ for device A. The black markers indicate experimentally-labeled data and the orange markers are the retrained and regularized CNN predictions. 
    \textbf{b.-d.} The product of the left and right differential conductance, $\GLL \cdot \GRR$, for varying $\VLM$ and $\VRM$. The measurements were performed at $\VABS$ values of \SI{44}{\milli\volt}, \SI{62}{\milli\volt} and \SI{135}{\milli\volt} respectively.
    \textbf{e.-g.} Non-local conductance, $\GRL = dI_\mathrm{R}/dV_\mathrm{L}$ for varying $\VLM$ and $\VRM$. The data was measured simultaneously with panels b-d.
    }\label{fig: restraints}
\end{figure}

As discussed in \cref{app:training}, the CNN was mainly trained on an effective model for the two QDs, that integrates out the ABS coupling them. When the QD-ABS coupling becomes on the order of the ABS energy, this model does not accurately describe the system anymore. In \cref{fig: restraints}a we see that the CNN predicts a sweet spot around $\VABS=\SI{45}{\milli\volt}$, while the labeled data indicates that $(t-\Delta)/(t+\Delta)\approx 1$. The weighted ratio approaches 1 when $\Delta \to  0$, as the numerator and denominator will be equal. From the corresponding CSD in \cref{fig: restraints}b we see that the correlated conductance vanishes, which points to weak interaction between the QDs. However, the corresponding non-local conductance in \cref{fig: restraints}e has both signs, which is usually a sign of a sweet spot. The features in panels b and e depend less on $\VRM$ than on $\VLM$, which suggests that the transport is dominated by the left QD. We interpret the presence of positive and negative non-local conductance as a sign of direct transport between the left QD and the ABS. From the tunnel spectroscopy in \cref{fig: 1}b, we see that the ABS at the lowest $\VABS$ values comes close to zero-energy at its minimum. This is consistent with our interpretation of an increased QD-ABS interaction at low $\VABS$ values.

In \cref{fig: restraints}c and f, we see both signs of non-local again, together with a cross in the correlated conductance. The non-local features depend mostly on $\VLM$, which suggests strong QD-ABS interaction as in panels b and e. It is hard to tell whether this is an actual sweet spot, and might be related to low Majorana polarization as detailed in Ref.~\cite{souto2023probing}. We note that the CNN predictions and experimentally extracted values of  $(t-\Delta)/(t+\Delta)$ agree at this $\VABS$ value.

Finally, we see that the labeled data becomes noisier around $\VABS=\SI{135}{\milli\volt}$ in \cref{fig: restraints}a. From panels d and g, we see that the QD-QD interaction is very weak. Accurately labeling the data becomes challenging for low interaction, as the peaks are hard to find due to low conductance. The CNN predictions are greater than the experimental labels here due to the regularization procedure. We conclude that the CNN performs well when the QD-ABS interaction is smaller than the ABS energy. Also, it does not perform well when the conductance becomes low due to weak QD-QD interaction.

\section{Regularization}\label{app:regularization} 
There are three cases that can lead to a vanishing $(t-\Delta)/(t+\Delta)$ ratio. In the first case, $(t+\Delta)$ becomes very large. The interaction strength is ultimately limited by the ABS energy and the superconducting parent gap. Second, $(t+\Delta)$ is finite, but $t \approx \Delta$, which is the desired sweet spot condition. In the third case, both $t$ and $\Delta$ become vanishingly small, but are not equal in size. While this should be mitigated by scaling $t-\Delta$ with $1/(t+\Delta)$, the CNN can incorrectly identify weak QD interaction as a sweet spot. To penalize low interaction, we add a correction to the objective function based on the mean conductance of a measurement. When the QDs have low interaction, the differential conductance will be negligible. First, we compute the average conductance of a charge stability diagram:
\begin{equation*}
    \braket{G} = \frac{1}{N}\sum_{i=1}^N \left( \frac{1}{M}\sum_{j=1}^M \GLL \cdot \GRR (\VLM^i, \VRM^j) \right)
\end{equation*}
Then we count the number of pixels of a measurement that have a conductance value greater than $\braket{G}$:
\begin{equation*}
    N_> = \sum_{i,j} ^{N,M} \Theta[\GLL \cdot \GRR (\VLM^i, \VRM^j)-\braket{G}]
\end{equation*}
Where $\Theta$ is the Heaviside function. Finally, we offset the objective function based on the ratio of above-average conductance pixels:
\begin{equation}
    f (\VABS) = \frac{\lvert t-\Delta \rvert}{t+\Delta} + \left( \frac{N}{2 N_>} \right)^3
\end{equation}\label{eq:regularization}
Where $N_>$ is limited by $0 \leq N_> \leq N^2$, assuming $M=N$. 
As $N_> \to \infty$, this correction disappears. As $N_> \to 0$, the correction becomes very large and a minimum of the objective function cannot be found. We use this regularization procedure for all the automated tuning runs.

\section{GRADIENT DESCENT ALGORITHM}\label{app:gradient} 
As detailed in \cref{eq:GD}, we use a gradient descent algorithm to find sweet spots where $t=\Delta$. While most gradient descent algorithms use the slope as a condition for convergence, we use the objective function. The reason is that we cannot compute the gradient at each given point, but need discrete points between which we can approximate the gradient using finite differences:
$$g_i =  \frac{f(\VABS)_i - f(\VABS)_{i-1}}{\VABS^i - \VABS^{i-1}}$$
We use a momentum term, $k$, to update the velocity with the average of the past 5 gradients in each step:
$$v^\mathrm{n+1} = k v^\mathrm{n} + (1-k) \braket{g}_n $$
Considering QD-based devices are mesoscopic in nature, predictions of the CNN at the same $\VABS$ can differ. This can result from gate jumps, gate hysteresis, off-centered measurements, etc. The momentum term helps to overcome local maxima of $f (\VABS)$ and find the global minimum.
We update $\VABS$ using the velocity scaled by a learning rate, $\eta$:
$$\VABS^{n+1}  = \VABS^{n}  - \eta v^\mathrm{n+1}$$
This procedure is repeated until the objective function is below the predefined tolerance, $\tau$:
$$f (\VABS) = \frac{\lvert t-\Delta \rvert}{t+\Delta} + \left( \frac{N}{2 N_>} \right)^3 <\tau $$
where the final term penalizes low QD-QD interaction, see \cref{app:regularization} for details. 

Considering the algorithm directly controls gate voltages on the device, we include some extra constraints. First, the minimal change in $\VABS$ is limited to $\SI{60}{\micro\volt}$ by our digital-analog converter resolution. Second, we limit the maximal change in $\VABS$ to $\SI{10}{\milli\volt}$. A larger change will shift the Coulomb resonance out of the measurement window. In each iteration, we center the measurements by taking linecuts of $\VLM$ and $\VRM$ at the extrema of a charge stability diagram, and interpolate their intersection.
Furthermore, we limit the algorithm with $\pm \SI{50}{\milli\volt}$ from each starting $\VABS$ to confine it to the known region. If the algorithm would propose a $\VABS$ across this boundary, we reverse the sign of the velocity. In the first iteration, we change $\VABS$ by a value drawn from a Gaussian distribution:
$$f (\Delta \VABS) = \frac{\SI{1.2}{\milli\volt}}{\sqrt{2 \pi}}e^{-\Delta \VABS^2/2}$$
While $g_i$ values are generally linear close to the sweet spot, they change sign at the sweet spot. If the CNN identifies a sign change of $t-\Delta$, we change the sign of the velocity -before updating it with the gradients- and reduce the learning rate by a factor of 2. We also increase the learning rate by 50\% every 5 steps.
\subsection*{Choice of learning rate}
For a single ABS in the infinite parent gap limit, we can write the elastic co-tunneling ($t$) and crossed Andreev reflection ($\Delta$) rates as \cite{liu2022tunable}:
\begin{equation*}
    t = \frac{\Gamma^{2} I_{0} \mu^{2}}{\left(\Gamma^{2} + \mu^{2}\right)^{2}} \ \ \ 
    \Delta = \frac{\Gamma^{4} I_{0}}{\Gamma^{4} + 2 \Gamma^{2} \mu^{2} + \mu^{4}}
\end{equation*}
Where $\Gamma$ is the induced gap of the ABS, $\mu$ is its electrochemical potential and $I_0$ is a proportionality constant. We can then write the sum and differences of $t$ and $\Delta$, as well as their ratio as:
\begin{equation}
    \begin{aligned}
    t+\Delta &=  \frac{\Gamma^{2} I_{0}}{\Gamma^{2} + \mu^{2}} \\
        t-\Delta &= \frac{\Gamma^{2} I_{0} \left(- \Gamma^{2} + \mu^{2}\right)}{\Gamma^{4} + 2 \Gamma^{2} \mu^{2} + \mu^{4}} \\
         \Lambda &=  \frac{t-\Delta}{t+\Delta}  =\frac{\mu^{2} - \Gamma^{2} }{\mu^{2} + \Gamma^{2}}
    \end{aligned}
\end{equation}
At $\mu = \pm \Gamma$, we obtain $t=\Delta$ and calculate the derivative:
$$\frac{d\Lambda}{d \mu} = \frac{4\mu \Gamma^2}{(\mu^2+\Gamma^2)^2}$$
\begin{equation}
    \left. \frac{d\Lambda}{d \mu} \right|_{\mu= \pm \Gamma} = \frac{1}{\Gamma}
\end{equation}
In a device, we are tuning $\mu$ indirectly using a gate voltage $\VABS$. We can write the slope in terms of $\VABS$ using:
\begin{align}
    \left. \frac{d\Lambda}{d \VABS} \right|_{\text{sweet spot}} &= \left. \frac{d\Lambda}{d \mu} \right|_{\mu= \pm \Gamma}  \frac{d\mu}{d\VABS} \\ &= \frac{1}{\Gamma}\frac{d\mu}{d\VABS} = \frac{\alpha}{\Gamma}
\end{align}
Where $\alpha$ is the lever arm. Bottou et al recommend a learning rate which matches the slope at the objective function minimum~\cite{bottou2012stochastic}. Using a lever arm $\alpha = 0.05e$ and an induced gap $\Gamma = \SI{100}{\micro\electronvolt}$, we obtain $\eta_\mathrm{opt} = \Gamma/\alpha = \SI{2e-3}{\volt}$, which is 10 times larger than the learning rate used in the experiments. We lower the learning rate to prevent large changes of $\VABS$, as cross-capacitance can shift the Coulomb resonances of the QDs.

We also note that we chose a relatively low learning rate, $\eta$, compared to the optimal value (see \cref{app:gradient} for details). This $\eta$ was chosen to prevent large changes in $\VABS$, which also affects the electrochemical potential of the QDs. The centering of the QD measurements is important for having accurate CNN predictions. This can be improved by using "virtual gates", which are linear combinations of  $\VLM$, $\VRM$ and $\VABS$ that compensate for cross-capacitance (see appendix B of Ref.~\cite{van2024charge}). Additionally, the CNN could be trained on more off-centered theory data. 


\subsection*{Choice of tolerance}
At $t=\Delta$, the QD levels intersect perfectly. As $t$ and $\Delta$ move away from the sweet spot, an avoided crossing of magnitude $D= \sqrt{8 \lvert t^2-\Delta^2 \rvert}$ opens between the two hyperbolas (see page 4 of the supplementary materials of~\cite{ten2024two}). If this magnitude is smaller than the broadening of conductance, an avoided crossing cannot be observed. We can define a tolerance based on broadening and the expected interaction strength $t+\Delta$. First, we rewrite the distance between parabola's as:
\begin{equation*}
    D = 2(t+\Delta)\sqrt{2 \frac{\lvert t-\Delta \rvert}{t+\Delta}}
\end{equation*}
If we demand that the distance is below thermal broadening, $D < 3.5 k_\mathrm{B}T$, we can write:
\begin{equation*}
\frac{\lvert t-\Delta \rvert}{t+\Delta} < \frac{1}{8}\left( \frac{3.5 k_\mathrm{B}T} {t+\Delta} \right)^2= \tau
\end{equation*}
Using a temperature of $T=\SI{30}{\milli\kelvin}$ and an interaction strength $t+\Delta = \SI{80}{\micro\electronvolt}$ we obtain a tolerance $\tau = \num{1.6e-3}$. While this quantity is smaller than the value we used in experiments, we note that our conductance broadening is significantly larger than the $\SI{9}{\micro\electronvolt}$ we can expect from thermal processes. Substituting the thermal term by a broadening of \SI{20}{\micro\electronvolt} yields a tolerance $\tau = \num{1e-2}$, which is the value we used in the experiments. Therefore we recommend using a tolerance:
\begin{equation*}
    \tau = \frac{1}{32}\left( \frac{\gamma} {\Delta^*} \right)^2
\end{equation*}
Where $\Delta^*$ is the desired gap at the sweet spot and $\gamma$ is the linewidth of the Coulomb resonances observed from bias spectroscopy. We note that in principle this allows us to select sweet spots based on their gap by choosing $\tau$. This is seen in \cref{fig: 5}b, where the algorithm converges on the weaker sweet spot at $\VABS = \SI{99.5}{\milli\volt}$ with a higher tolerance than on the stronger one at $\VABS = \SI{123.5}{\milli\volt}$. Although this implies that we can reject weakly-interacting sweet spots, they are still minima of the objective function. A possible improvement of the algorithm could be lowering the tunnel barriers when the objective function does not converge below tolerance, or penalizing the objective function for weak sweet spots, as in \cref{fig: 6}.

\subsection*{Pseudo code}
In \cref{alg:1} we show the pseudo code for the automated tuning algorithm.
\begin{algorithm}[H]
\caption{Sweet Spot Optimization Algorithm}
\begin{algorithmic}[1]
\Require $V\_ABS\_parameter$, $learning\_rate$, $epochs$, $min\_step\_size$, $tolerance$, $N$, $momentum$, $damping\_term$, $max\_step\_size$, $V\_ABS\_min$, $V\_ABS\_max$

\State Assert $min\_step\_size > 1e-6$


\For{$epoch = 1$ \textbf{to} $epochs$}
    \If{$epoch == 1$}
        \State $V\_ABS \gets$ handle\_first\_epoch($V\_ABS$, $min\_step\_size$)
    \Else
        \If{$epoch \mod 5 == 0$}
            \State $learning\_rate \gets learning\_rate \times 1.5$
        \EndIf
        \State Compute gradients
        \State Update $V\_ABS$ and $v$ using gradients
    \EndIf

    \State Set gate value using $V\_ABS\_parameter(V\_ABS)$
    \State Precenter measurement
    \State $dbx \gets measure\_func(measure\_func\_kwargs)$
    \State Update $delta\_min\_t\_values$, $obj\_fun\_values$, $V\_ABSs$, $dbxs$
    \If{Overshoot detected}
        \State Reverse $v$, reduce $learning\_rate$, log overshoot
    \EndIf

    \If{$\lvert current\_obj\_val \rvert < tolerance$}
        \State Log convergence, break
    \EndIf
\EndFor

\If{Non-convergence after all epochs}
    \State Log non-convergence details
\EndIf

\State Close log
\Return $V\_ABSs$, $obj\_fun\_values$, $current\_obj\_val$, $epoch$
\end{algorithmic}\label{alg:1}
\end{algorithm}

\section{VALIDATION OF POOR MAN'S MAJORANA ZERO MODES IN SWEET SPOTS}
\label{app:validation}
The histogram in \cref{fig: 5}b indicates that a large majority of runs end at $\VABS \approx \SI{90.5}{\milli\volt}$ and $\VABS \approx \SI{123.5}{\milli\volt}$. Based on the charge stability diagrams in panel c, the classifier identifies sweet spots at those $\VABS$ values, by predicting $\frac{|t-\Delta|}{t+\Delta}$ to fall below a set tolerance. The CSDs here form a cross-shape, which signals a zero-energy state that does not split upon tuning a single QD plunger gate. The non-local conductance in \cref{fig: 5}d  having both positive and negative sign indicates the presence of finite ECT and CAR rates, a condition for the emergence of Poor Man’s Majorana zero modes in a two-site Kitaev chain. 

While \cref{fig: 5}c,d  indicate a sweet spot at which Poor Man’s Majorana zero modes emerge, it does not reveal the size of the gap to the first excited state. 
\begin{figure}[t!]
    \centering
    \includegraphics[width=\linewidth]{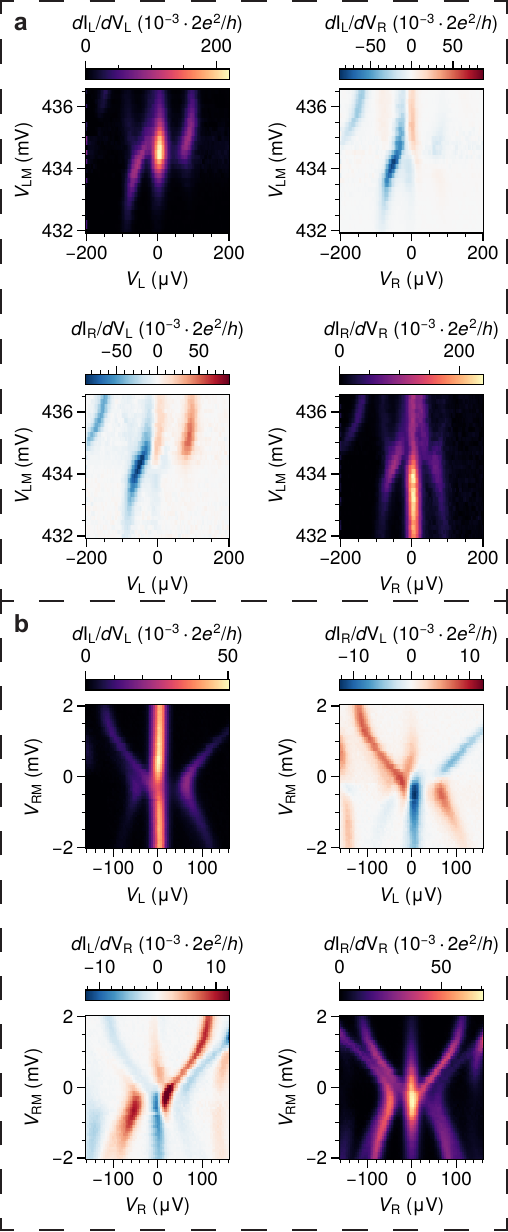}
    \caption{\textbf{zero-energy state stability for device A at the sweet spots identified in \cref{fig: 5} at zero magnetic field.} 
    \textbf{a.} Full conductance matrix for varying $\VLM$ and voltage bias $V_\mathrm{L/R}$ at $\VABS = \SI{95}{\milli\volt}$.
    \textbf{b.} Full conductance matrix for varying $\VRM$ and $V_\mathrm{L/R}$ at $\VABS = \SI{120}{\milli\volt}$.
    }\label{fig: validation_nanowire}
\end{figure}
In \cref{fig: validation_nanowire}a, we show the local and non-local conductance measured on device A on varying $\VLM$, while the right QD is kept on resonance. We observe a stable zero-energy state, combined with dispersing excited states. Similarly, in panel b we show that the sweet spot at $\SI{120}{\milli\volt}$ displays a stable zero-bias peak as well, upon varying $\VRM$, with a gap to excited states. As detailed in Refs.~\cite{zatelli2024robust, dvir2023realization, ten2024two}, this is the robustness against local perturbation that defines Poor Man’s Majorana zero modes in a two-site Kitaev chain, and can be used to validate sweet spots identified by the automated tuning algorithm.

The gate voltages at which these sweet spots are found differ slightly from the ones in \cref{fig: 5}. As these spectroscopic measurements were performed before the automated tuning runs, we attribute the difference to drifts in gate voltage. We identify these as the same sweet spots shown in \cref{fig: 5}, as there is only a single ABS in this gate range, as seen from \cref{fig: 1}b, which coincides with two expected sweet spots~\cite{zatelli2024robust}.

\begin{figure}[t!]
    \centering
    \includegraphics[width=\linewidth]{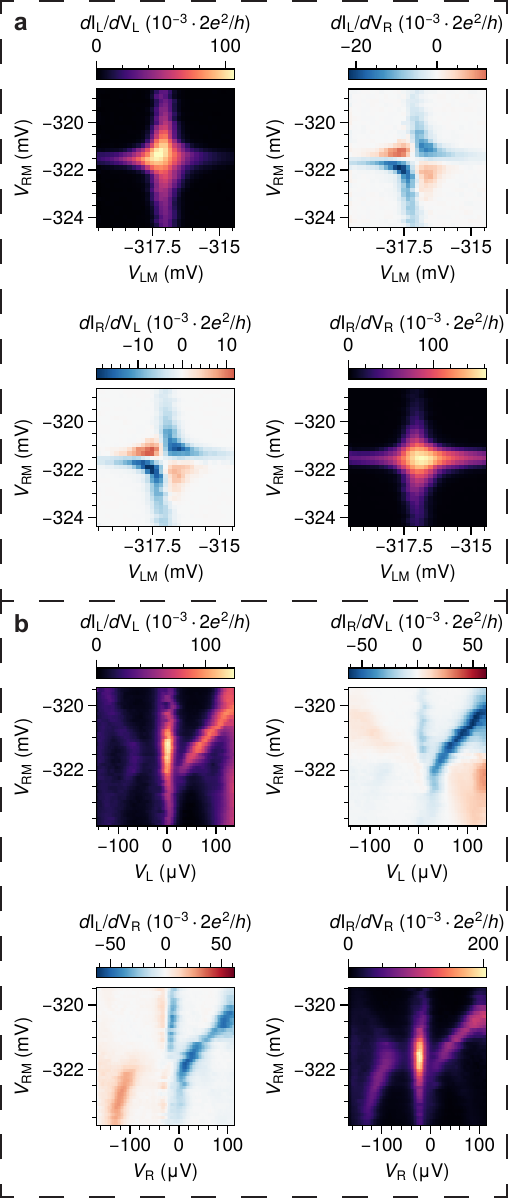}
    \caption{\textbf{zero-energy state stability for device B at $\VABS = \SI{-623.3}{\milli\volt}$ at zero magnetic field.} 
    \textbf{a.} Full conductance
matrix of a crossing in the CSD, when $t\approx\Delta$, for varying $\VLM$ and $\VRM$.
    \textbf{b.} Full conductance matrix for varying $\VRM$ and voltage bias $V_\mathrm{L/R}$. 
    }\label{fig: validation_2DEG}
\end{figure}
In \cref{fig: validation_2DEG}a, we show a CSD of device B in the sweet spot used for the training data. We note that this measurement was taken at $\VABS = \SI{-622.2}{\milli\volt}$. This voltage also shows a small difference when compared to the gate voltage at which $\frac{|t-\Delta|}{t+\Delta}=0$ in \cref{fig: 3}a,b. Similarly to the previous case the sweet spot was not characterized at the same time as the collection of the labeled data. As such, we attribute the difference to the drifts in gate voltages.
In \cref{fig: validation_2DEG}b, we show the local and non-local conductance measured on device upon varying $\VRM$, while the left QD is kept on resonance. Similar to device A, we see a stable zero-bias peak with a gap to excited states. 

The Poor Man's Majorana confirmation step via spectroscopic measurements (\cref{fig: validation_nanowire} and \cref{fig: validation_2DEG}), as shown in this appendix, can be always performed once a candidate sweet spot is identified by the algorithm

\section{UNCONVERGED RUN EXAMPLE}\label{app:unconverged example} 
\begin{figure*}[ht!]
    \centering
    \includegraphics[width=\linewidth]{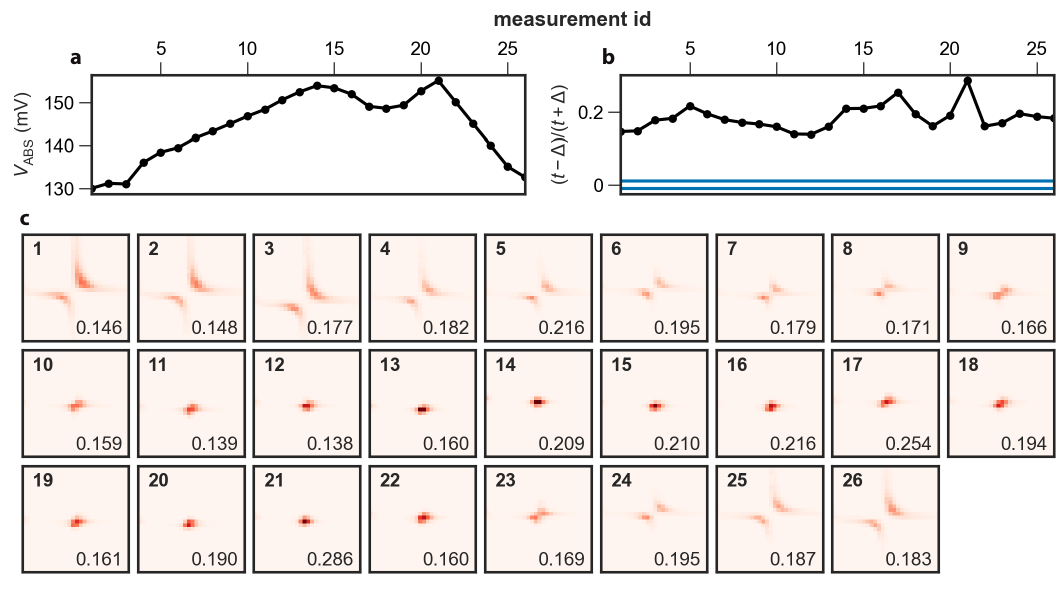}\caption{\textbf{Example of an unconverged automated tuning run.} 
    \textbf{a.} The hybrid plunger gate voltage, $\VABS$, for each measurement.
    \textbf{b.} Prediction of the weighted ratio of elastic co-tunneling and crossed-Andreev reflection, $(t-\Delta)/(t+\Delta)$, for each measurement. The horizontal, blue lines indicate the tolerance set before the run.
    \textbf{c.} The product of the left and right differential conductance, $\GLL \cdot \GRR$, for varying $\VLM$ and $\VRM$ for subsequent measurements. The measurement ids are indicated by the bold text in the top left corner. The predicted value of $(t-\Delta)/(t+\Delta)$ is indicated in the bottom right corner.
    }\label{fig: bad example}
\end{figure*}

In \cref{fig: bad example}a, we show the $\VABS$ values visited by a run that did not converge. Just as in \cref{fig: 4}, the run started from $\VABS = \SI{130}{\milli\volt}$. For the first 14 measurements, the algorithm visits increasing values of $\VABS$. We see from \cref{fig: bad example}b that $(t-\Delta)/(t+\Delta)$ does not change significantly in this gate range. From the charge stability diagrams in \cref{fig: bad example}c, we see that the interaction between the QDs becomes weaker for increasing $\VABS$. Eventually, the algorithm reverses direction and returns to the starting $\VABS = \SI{130}{\milli\volt}$. As we limit the number of measurements per run to 26, the algorithm stops here. This unconverged run highlights that the CNN does not predict $(t-\Delta)/(t+\Delta)$ accurately if $t+\Delta$ is small, as can be seen from the superimposed CNN predictions in the charge stability diagrams. We attribute this to the bias toward strong interaction in our theoretical and experimental training set. We note that this run ended with a velocity in the right direction and might have converged given more steps.

\section{A SPINFUL HAMILTONIAN}\label{app:spinful hamiltonian} 
\begin{figure}[ht!]
\centering
\includegraphics[width=\linewidth]{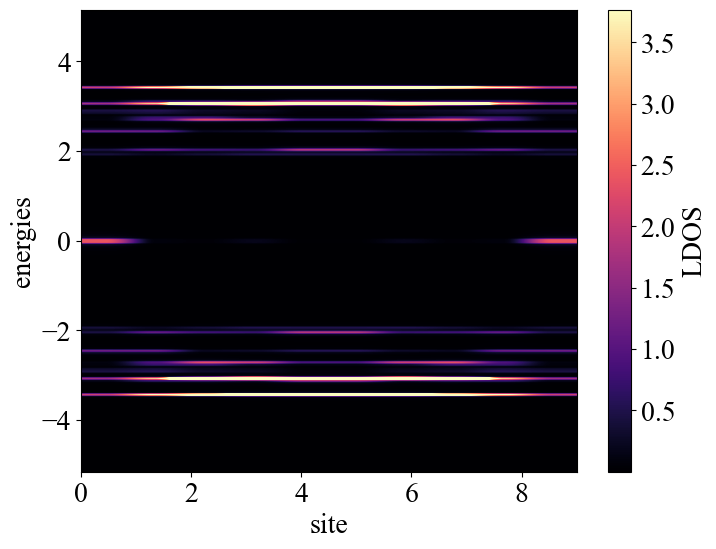}
\caption{
Example of the LDOS of a 10-site chain that shows zero energy edge modes. 
}
\label{fig: LDOS}
\end{figure}

\begin{figure}[ht!]
\centering
\includegraphics[width=\linewidth]{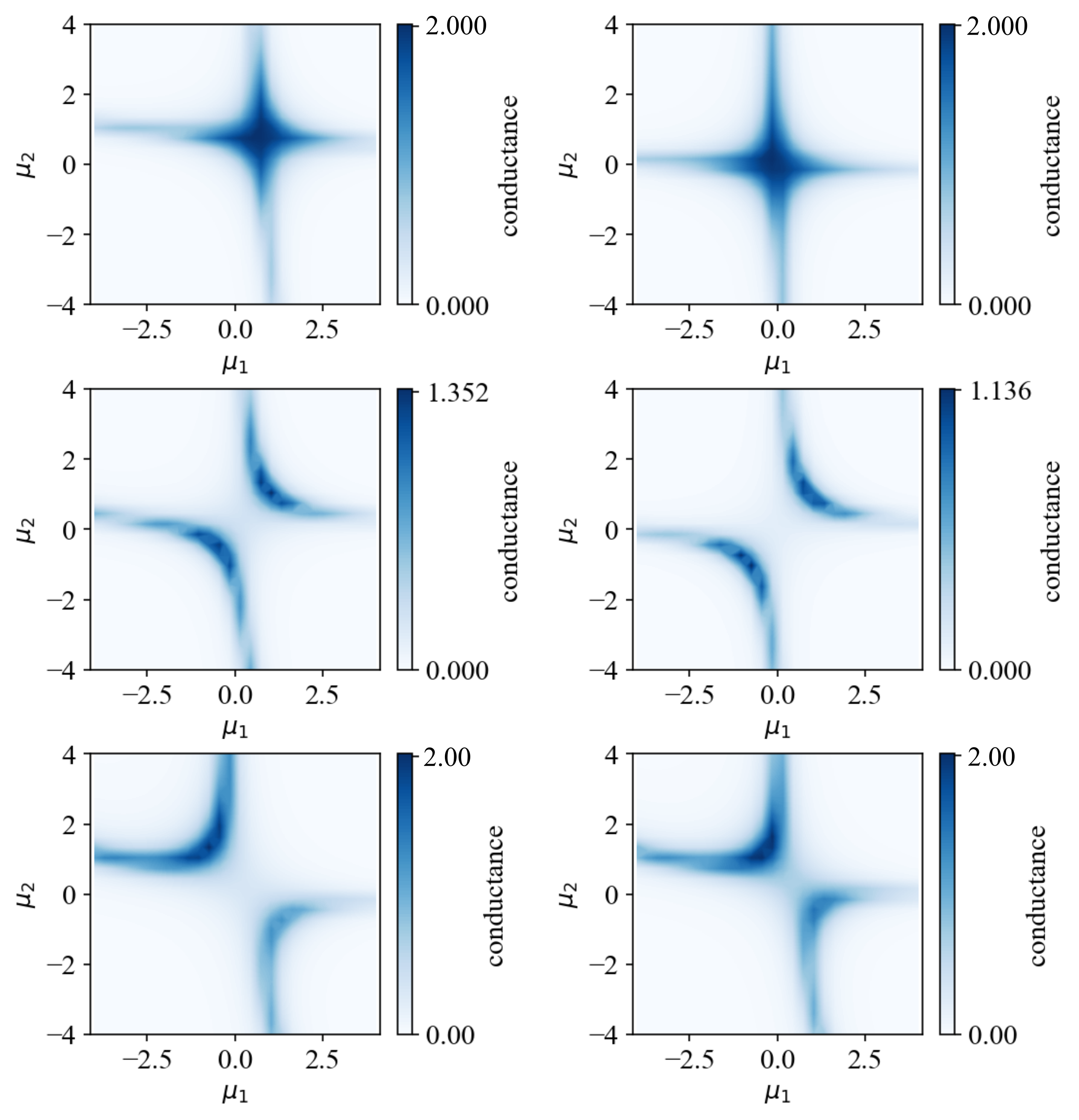}
\caption{
Simulation of conductance with Hamiltonian from Eq.\ref{eq: spinful} and S-matrix formalism using the pyqula library \cite{Lado_pyqula_2025}.
}
\label{fig: conductance}
\end{figure}

Here, we introduce a Hamiltonian that has all the required ingredients to study Majorana bound states and topological superconductivity.
This Hamiltonian describes a spinful quantum dot system including superconductivity, spin-orbit coupling, and Zeeman splitting and can be written as
\begin{equation}\label{eq: spinful}
H = H_0 + H_{\text{SOC}} + H_Z + H_{\text{s-wave}}.
\end{equation}
The tight-binding Hamiltonian $H_0$ describes the movement of electrons in the system that determines the basic electronic band structure. This term is defined as
\begin{equation}
H_0 = \sum_{i,\sigma} \mu_i c_{i\sigma}^\dagger c_{i\sigma} - t \sum_{\langle i,j \rangle, \sigma} (c_{i\sigma}^\dagger c_{j\sigma} + h.c.)
\end{equation}
where $\mu_i$ is the onsite chemical potential at site $i$, $t$ is the hopping amplitude between neighboring sites $\langle i, j \rangle$, and $c_{i\sigma}^\dagger$ and $c_{i\sigma}$ are the electron creation and annihilation operators at site $i$ with spin $\sigma$.
The Rashba spin-orbit coupling (SOC) Hamiltonian $H_{\text{SOC}}$ is given as
\begin{equation}
H_{\text{SOC}} = i \lambda_R \sum_{\langle i,j \rangle} \left( c_{i\uparrow}^\dagger c_{j\downarrow} - c_{i\downarrow}^\dagger c_{j\uparrow} \right) + h.c.
\end{equation}
where $\lambda_R$ is the Rashba SOC strength. This term breaks the spin degeneracy and allows spin flips during hopping.
The Zeeman term $H_Z$
\begin{equation}
H_Z = B_z \sum_i \left( c_{i\uparrow}^\dagger c_{i\uparrow} - c_{i\downarrow}^\dagger c_{i\downarrow} \right)
\end{equation}
breaks spin symmetry and induces energy splitting between spin-up and spin-down states. $B_z$ represents the Zeeman energy proportional to the applied magnetic field.
This term removes the spin degeneracy and shifts the energy levels of spin-up and spin-down electrons.
The superconducting s-wave pairing term $H_{\text{s-wave}}$ is given by
\begin{equation}
H_{\text{s-wave}} = \sum_i \Delta_s \left( c_{i\uparrow}^\dagger c_{i\downarrow}^\dagger + h.c. \right)
\end{equation}
where $\Delta_s$ is the superconducting pairing. All terms together lead to the full spinful Hamiltonian 
\begin{equation}
    \begin{split}
        H = & \sum_{i,\sigma} \mu_i c_{i\sigma}^\dagger c_{i\sigma} - t \sum_{\langle i,j \rangle, \sigma} (c_{i\sigma}^\dagger c_{j\sigma} + h.c.) \nonumber\\
         & + i \lambda_R \sum_{\langle i,j \rangle} \left( c_{i\uparrow}^\dagger c_{j\downarrow} - c_{i\downarrow}^\dagger c_{j\uparrow} \right) \nonumber \\
        & + B_z \sum_i (c_{i\uparrow}^\dagger c_{i\uparrow} - c_{i\downarrow}^\dagger c_{i\downarrow}) \nonumber  + \sum_i \Delta_s \left( c_{i\uparrow}^\dagger c_{i\downarrow}^\dagger + h.c. \right).
    \end{split}
\end{equation}
By adjusting the parameters $\lambda_R$, $B_z$, and $\Delta_s$ we can tune the system in regions where the model can host Majorana bound states or Poor man's Majoranas in the case of a 2-dot system. 
Compared to the Kitaev model, this Hamiltonian explicitly includes the spin degrees of freedom. 
In the limit of a strong Zeeman field $B_Z >> t, \lambda_R$, $B_z$, $\Delta_s$, the spinful Hamiltonian approximates the Kitaev model. 
In Fig. \ref{fig: LDOS}, we compute the local density of states (LDOS) to show the appearance of zero energy edge modes.
This model allows us to compute the conductance for different regimes. In Fig.\ref{fig: conductance}, we show that we obtain the same avoided crossings as for the Kitaev model, as well as the features for the Poor Man's Majorana sweet spot.

To train and test a CNN with the spinful model, we create a dataset of 1000 examples and split it into training and testing (80/20). We simulate the data, including the three parameter regimes shown in Fig.\ref{fig: conductance}, namely the Marjorana sweet spot and both avoided crossings (corresponding to the three regimes in the Kitaev chain model).
We then train the CNN on these simulations to predict which regime the system is in, based on the avoided crossings of the conductance simulations. 
On the theory data, the CNN achieved 99.1$\%$ accuracy for classification on the test set.
Next, we take the 2DEG data (Fig.\ref{fig: 3} f of the manuscript) and use the same CNN to classify between the two regimes of avoided crossings. 
In this case, we obtain 90.7$\%$ (68 out of 75) accuracy on the experimental data, which is in comparison to the CNN prediction of previous work with the Kitaev model for experimental data 92-94$\%$ accuracy \cite{Koch2023}.

\clearpage

\bibliography{Main}
\end{document}